\documentclass[]{pasj01}
\usepackage{natbib,color}

\begin{document} 
\Received{}
\Accepted{}

\title{An optically-selected cluster catalog at redshift $0.1<z<1.1$ from the
  Hyper Suprime-Cam Subaru Strategic Program S16A data}

\author{Masamune \textsc{Oguri}\altaffilmark{1,2,3}}%
\author{Yen-Ting \textsc{Lin}\altaffilmark{4}}
\author{Sheng-Chieh \textsc{Lin}\altaffilmark{4,5}}
\author{\\Atsushi J. \textsc{Nishizawa}\altaffilmark{6,7}}
\author{Anupreeta \textsc{More}\altaffilmark{3}}
\author{Surhud \textsc{More}\altaffilmark{3}}
\author{\\Bau-Ching \textsc{Hsieh}\altaffilmark{4}}
\author{Elinor \textsc{Medezinski}\altaffilmark{8}}
\author{Hironao \textsc{Miyatake}\altaffilmark{9,3}}
\author{\\Hung-Yu \textsc{Jian}\altaffilmark{4}}
\author{Lihwai \textsc{Lin}\altaffilmark{4}}
\author{Masahiro \textsc{Takada}\altaffilmark{3}}
\author{Nobuhiro \textsc{Okabe}\altaffilmark{10,11}}
\author{\\Joshua S. \textsc{Speagle}\altaffilmark{12,3}}
\author{Jean \textsc{Coupon}\altaffilmark{13}}
\author{Alexie \textsc{Leauthaud}\altaffilmark{14,3}}
\author{\\Robert H. \textsc{Lupton}\altaffilmark{8}}
\author{Satoshi \textsc{Miyazaki}\altaffilmark{15,16}}
\author{Paul A. \textsc{Price}\altaffilmark{8}}
\author{\\Masayuki \textsc{Tanaka}\altaffilmark{15}}
\author{I-Non \textsc{Chiu}\altaffilmark{4}}
\author{Yutaka \textsc{Komiyama}\altaffilmark{15,16}}
\author{\\Yuki \textsc{Okura}\altaffilmark{17,18}}
\author{Manobu M. \textsc{Tanaka}\altaffilmark{19}}
\author{Tomonori \textsc{Usuda}\altaffilmark{15,16}}

\altaffiltext{1}{Research Center for the Early Universe, University of Tokyo, Tokyo 113-0033, Japan}
\altaffiltext{2}{Department of Physics, University of Tokyo, Tokyo 113-0033, Japan}
\altaffiltext{3}{Kavli Institute for the Physics and Mathematics of the Universe (Kavli IPMU, WPI), University of Tokyo, Chiba 277-8582, Japan}
\altaffiltext{4}{Institute of Astronomy and Astrophysics, Academia Sinica, P.O. Box 23-141, Taipei 10617, Taiwan}
\altaffiltext{5}{Department of Physics, National Taiwan University, 10617 Taipei, Taiwan}
\altaffiltext{6}{Institute for Advanced Research, Nagoya University, Aichi 464-8601, Japan}
\altaffiltext{7}{Department of physics, Nagoya University, Aichi 464-8602, Japan}
\altaffiltext{8}{Department of Astrophysical Sciences, Princeton University, Princeton, NJ 08544, USA}
\altaffiltext{9}{Jet Propulsion Laboratory, California Institute of Technology, Pasadena, CA 91109, USA}
\altaffiltext{10}{Department of Physical Science, Hiroshima University, 1-3-1 Kagamiyama, Higashi-Hiroshima, Hiroshima 739-8526, Japan}
\altaffiltext{11}{Hiroshima Astrophysical Science Center, Hiroshima University, Higashi-Hiroshima, 1-3-1 Kagamiyama 739-8526, Japan}
\altaffiltext{12}{Department of Astronomy, Harvard University, Cambridge, MA 02138, USA}
\altaffiltext{13}{Astronomical Observatory of the University of Geneva, ch. d’Ecogia 16, CH-1290 Versoix, Switzerland}
\altaffiltext{14}{Department of Astronomy and Astrophysics, University of California Santa Cruz, Santa Cruz, CA 95064, USA}
\altaffiltext{15}{National Astronomical Observatory of Japan, Mitaka, Tokyo 181-8588, Japan}
\altaffiltext{16}{Department of Astronomy, School of Science, Graduate University for Advanced Studies, Mitaka, Tokyo 181-8588, Japan}
\altaffiltext{17}{RIKEN Nishina Center, 2-1 Hirosawa, Wako, Saitama 351-0198, Japan}
\altaffiltext{18}{RIKEN-BNL Research Center, Department of Physics, Brookhaven National Laboratory, Bldg. 510, Upton, NY, 11792, USA}
\altaffiltext{19}{High Energy Accelerator Research Organization (KEK), Institute of Particle and Nuclear Studies, 1-1 Oho, Tsukuba 305-0801}

\email{masamune.oguri@ipmu.jp}





\maketitle

\begin{abstract}
We present an optically-selected cluster catalog from the Hyper
Suprime-Cam (HSC) Subaru Strategic Program. The HSC images are
sufficiently deep to detect cluster member galaxies down to
$M_*\sim 10^{10.2}M_\odot$  even at $z\sim 1$, 
allowing a reliable cluster detection at such high redshifts. We apply
the CAMIRA algorithm to the HSC Wide S16A dataset covering $\sim
232$~deg$^2$ to construct a catalog of 1921 clusters at redshift
$0.1<z<1.1$ and richness $\hat{N}_{\rm mem}>15$ that roughly
corresponds to $M_{\rm 200m}\gtrsim 10^{14}h^{-1}M_\odot$. We confirm
good cluster photometric redshift performance, with the bias and
scatter in  $\Delta z/(1+z)$ being better than 0.005 and 0.01 over
most of the redshift range, respectively. We compare our cluster
catalog with large X-ray cluster catalogs from XXL and XMM-LSS surveys
and find good correlation between richness and X-ray properties. We
also study  the miscentering effect from the distribution of offsets
between optical and X-ray cluster centers. We confirm the high ($>0.9$)
completeness and purity for high mass clusters by analyzing mock
galaxy catalogs.
\end{abstract}

\section{Introduction}

Clusters of galaxies are dominated by dark matter, which makes them
a useful site for cosmological studies. For example, the abundance of
massive clusters and its time evolution, which can well be predicted
by $N$-body simulations, are known to be a sensitive probe of
cosmological parameters \citep[e.g.,][]{allen11,weinberg13}. Detailed
studies of dark matter distributions in clusters basically support the
so-called cold dark matter paradigm at the non-linear scale
\citep[e.g,][]{oguri10,oguri12,okabe10,okabe13,umetsu11,umetsu16,niikura15}. 
In addition, clusters of galaxies
play an essential role in understanding galaxy formation physics given
a possible large environmental effect on galaxy formation and evolution
\citep[e.g.,][]{renzini06,kravtsov12}.  

Clusters can be identified in various wavelengths, including optical,
X-ray, and radio/mm/submm via the Sunyaev-Zel'dovich
\citep[SZ;][]{sunyaev72} effect. While there are advantages and
disadvantages for each method, the recent development of wide-field
optical imaging surveys makes surveys of clusters in optical
particularly powerful, because they take wide-field images with
multi-bands, which is crucial both in selecting clusters of galaxies
efficiently from the enhancement of galaxy number densities as well as
deriving photometric redshifts of clusters
\citep[e.g.,][]{gladders00}. Indeed, large samples of
optically-selected clusters have been constructed in Sloan Digital Sky
Survey \citep{koester07,hao10,szabo11,wen12,rykoff14,oguri14}, the
Red-Sequence Cluster Survey \citep{gladders05}, the
Canada-France-Hawaii-Telescope Legacy Survey
\citep{milkeraitis10,ford14,licitra16}, the Blanco Cosmology Survey
\citep{bleem15}, and the Dark Energy Survey Science Verification Data
\citep[DES;][]{rykoff16}. Because of the depth and wavelength coverage
of these optical surveys, the redshift range of most of these clusters
are restricted to $z\lesssim 0.9$ at most. Attempts to find clusters
at higher redshifts have also been made using infrared data, although
with generally smaller area
\citep[e.g.,][]{goto08,wilson09,andreon09,gettings12,rettura14}. 

The Hyper Suprime-Cam \citep[HSC;][]{miyazaki12,miyazaki15} is a new
wide-field optical imager installed on the Subaru 8.2-meter telescope.
The HSC Subaru Strategic Program (hereafter the HSC Survey) is
an ongoing wide-field optical imaging survey \citep{aihara17}.
It consists of three layers (Wide, Deep, and Ultradeep), and the Wide
layer is planned to observe the total sky area of $\sim1400$~deg$^2$
with five broadband filters ($grizy$). With its unique combination of
area and depth ($i_{\rm lim}\sim 26$ at $5\sigma$), the HSC Wide layer is
expected to revolutionize the optically-selected cluster search by
constructing a large sample of massive clusters at $z\sim 1$ and beyond.  

In this paper, we construct a new optically-selected cluster catalog
from the first two years of observation of the HSC Survey, using the
CAMIRA (Cluster finding Algorithm based on Multi-band Identification of
Red-sequence gAlaxies) algorithm \citep{oguri14}. We construct a
catalog of 1921 clusters with richness $\hat{N}_{\rm mem}>15$ and the
redshift range $0.1<z<1.1$. The catalog is compared with spectroscopic
and X-ray data as well as mock galaxy catalogs to check its validity.  

This paper is organized as follows. In Section~\ref{sec:data} we
describe the HSC data and the cluster finding algorithm used in the
paper. Section~\ref{sec:catalog} presents the cluster catalog and
discusses the accuracy of photometric redshifts of the clusters.
We compare the HSC cluster catalog with the SDSS CAMIRA cluster
catalog in Section~\ref{sec:sdss}, and X-ray cluster catalogs in
Section~\ref{sec:xray}. We present an analysis of mock galaxy catalogs
in Section~\ref{sec:mock}. We summarize our results in
Section~\ref{sec:summary}. Throughout the paper we assume a flat
$\Lambda$-dominated cold dark matter model with $\Omega_M=0.28$,
$\Omega_\Lambda=0.72$, $h=0.7$, $\Omega_b=0.044$, $n_s=0.96$, and
$\sigma_8=0.82$. Magnitudes in this paper are in the AB magnitude
system, and are corrected for Galactic extinction 
\citep{schlegel98}. 

\section{Data and Method}\label{sec:data}

\subsection{HSC Wide Data}\label{sec:hscdata}

In this paper we use the S16A internal data release of the HSC Survey,
which was released in 2016 August. The S16A release contains imaging
data taken between 2014 March and 2016 April, containing 174~deg$^2$
of the HSC Wide data taken in all five broadbands at full depth.
The area that covers all five broadbands but with non full depth
exceeds 200~deg$^2$. 
The HSC data are reduced with the HSC Pipeline, hscPipe
\citep{bosch17}, which is based on the Large Synoptic Survey Telescope 
pipeline \citep{ivezic08,axelrod15,juric15}. The Pan-STARRS1 data
\citep{tonry12,schlafly12,magnier13} are also used for astrometric and
photometric calibrations. 

We create an input galaxy catalog from the HSC database.
We select galaxies that were observed in all five broadbands, by imposing
cuts in the number of visits for each object using {\tt countinputs}
parameter, which indicates the number of images used to create a coadd
image for each galaxy. Specifically we set {\tt countinputs}$\geq 2$
for $gr$-bands and {\tt countinputs}$\geq 4$ for $izy$-bands. While
this condition is a more relaxed condition than the nominal definition
of the full depth for the HSC Wide Survey ({\tt countinputs}$= 4$ for
$gr$-bands and {\tt countinputs}$=6$ for $izy$-bands), we adopt this
condition to avoid gaps in the galaxy distribution due to CCD gaps. We
only use galaxies with $z$-band {\tt cmodel} magnitude brighter than
$z=24$, its error smaller than $0.1$, and $i$-band star-galaxy separation
parameter {\tt classification\_extendedness}$=1$.
We use $z$-band for the detection magnitude, rather than $i$-band as
used in the SDSS CAMIRA cluster catalog, as it is better suited for
galaxies at redshift around unity. The limiting magnitude of $z=24$
corresponds to more than $10\sigma$ detection significance for most of
the area of interest, and hence the detection completeness is very
close to unity. On the other hand, we use $i$-band star-galaxy
separation because $i$-band images are on average taken in better
seeing conditions with a median seeing size of $\sim 0\farcs6$ than the
other bands, in order to allow accurate galaxy shape measurements for
weak lensing analysis. In addition we place weak constraints on $ri$-bands as
$r<26.5$ and $i<26$ mainly to remove artifacts. We also remove
galaxies that can be affected by bad pixels or have poor photometric
measurements, by rejecting objects with any of the following flags in
any five broadbands;
{\tt flags\_pixel\_edge}, {\tt flags\_pixel\_cr\_center},
{\tt flags\_pixel\_interpolated\_center},  
{\tt cmodel\_flux\_flags}, and {\tt parent\_flux\_convolved\_2\_0\_flags}.
In addition we remove objects with the flag {\tt
  centroid\_sdss\_flags} in $riz$-bands \citep{aihara17}

We use {\tt cmodel} magnitudes, which are magnitudes derived from
light profile fitting, as ``total'' magnitudes of galaxies
\citep{bosch17}. However, we find that photometry with the current
version of hscPipe can be 
inaccurate in crowded regions such as cluster centers because of
failure of the deblender in such regions. More specifically, although
hscPipe successfully deblends galaxies in crowded regions, magnitudes
of deblended galaxies in highly crowded regions are sometimes offset
significantly. Given the critical importance of accurate colors for
cluster finding, in this paper we adopt the following hybrid approach
for galaxy photometry. We derive colors of individual galaxies using
aperture photometry on the `parent' (i.e., undeblended) image after
the Point Spread Function (PSF) sizes are matched between five
broadbands, because this aperture photometry provide accurate colors
even for galaxies in crowded regions. We use the target PSF size of
$1\farcs1$ and the aperture size of $1\farcs1$ in diameter ({\tt
  parent\_mag\_convolved\_2\_0}). As stated above, we use {\tt cmodel}
magnitudes, which are galaxy magnitudes from light profile fitting,
for total $z$-band magnitudes, and derive magnitudes in the other
bands using the colors from the aperture photometry on PSF-matched
images as described above in detail. 

The HSC database also provides a bright star mask. Masked areas are
dependent on magnitudes of stars, and are chosen conservatively. As a
result, about 10\% of the area fall in the bright star mask. Here we do
not apply the bright star mask to create our cluster catalog, although
we also provide an HSC Wide cluster catalog constructed with the
bright star mask in Appendix~1. 
The current version of the bright star mask contains an issue that
affects cluster finding at low redshifts, which is discussed in
Appendix~2. The input galaxy catalog contains 19,060,802 galaxies.  

\subsection{Spectroscopic Data}\label{sec:specgal}

\begin{figure}
 \begin{center}
  \includegraphics[width=8cm]{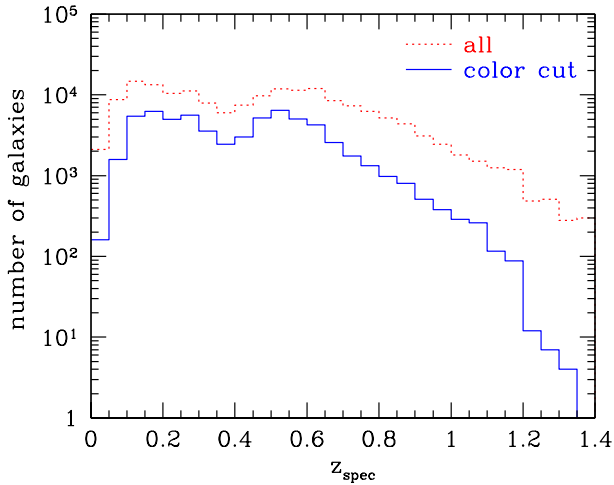} 
 \end{center}
\caption{The number distribution of spectroscopic galaxies in the
  HSC Wide Survey as a function of spectroscopic redshift. The dotted
  line shows the distribution for all the spectroscopic galaxies. The
  solid line shows the distribution after the color cuts to select
  red-sequence galaxies (see the text for details). }\label{fig:hist_specz} 
\end{figure}

As discussed in \cite{oguri14}, we need a calibration of red-sequence
galaxy colors with spectroscopic galaxies in order to improve the
accuracy of our algorithm. In fact, the HSC Survey overlaps with a
number of spectroscopic surveys including SDSS DR12 \citep{alam15},
DEEP2 DR4 \citep{newman13}, PRIMUS DR1 \citep{coil11}, 
VIPERS PDR1 \citep{garilli14}, VVDS \citep{lefevre13},
GAMA DR2 \citep{liske15}, WiggleZ DR1 \citep{drinkwater10},
zCOSMOS DR3 \citep{lilly09}, UDSz \citep{bradshaw13,mclure13},
3D-HST v4.1.5 \citep{momcheva16}, and FMOS-COSMOS v1.0 \citep{silverman15}.
We use these spectroscopic data for the calibration. 

First we cross match the spectroscopic catalogs with the HSC Wide S16A
galaxy catalog constructed in Section~\ref{sec:hscdata} to obtain a
sample of 171,255 galaxies with spectroscopic redshifts between
$z=0.01$ and 1.4 and the redshift error smaller than 0.01. Since we
are interested in red-sequence galaxies only, we apply additional
color cuts to remove obvious blue galaxies and to select galaxies used
for the calibration of the red-sequence colors. Specifically we adopt
the following color cuts 
\begin{equation}
  g-r>
\left\{
    \begin{array}{l}
      0.398+2.9z_{\rm spec}\;\;\;(z_{\rm spec} \leq 0.38), \\
      1.861-0.95z_{\rm spec}\;\;\;(0.38 < z_{\rm spec} ),
    \end{array}
  \right.
\label{eq:gr1}
\end{equation}
\begin{equation}
  r-i>
\left\{
    \begin{array}{l}
      0.2+0.8z_{\rm spec}\;\;\;(z_{\rm spec} \leq 0.369), \\
      -0.169+1.8z_{\rm spec}\;\;\;(0.369 < z_{\rm spec} \leq 0.75), \\
       1.481-0.4z_{\rm spec}\;\;\;(0.75 < z_{\rm spec} ),
    \end{array}
  \right.
\label{eq:ri1}
\end{equation}
\begin{equation}
  r-i<2-0.4z_{\rm spec},
\end{equation}
\begin{equation}
  i-z>
\left\{
    \begin{array}{l}
      0.122+0.3z_{\rm spec}\;\;\;(z_{\rm spec} \leq 0.76), \\
      -1.132+1.95z_{\rm spec}\;\;\;(0.76 < z_{\rm spec} \leq 0.96), \\
       0.164+0.6z_{\rm spec}\;\;\;(0.96 < z_{\rm spec} ),
    \end{array}
  \right.
\label{eq:iz1}
\end{equation}
\begin{equation}
  i-z<0.5+0.5z_{\rm spec},
\end{equation}
where $z_{\rm spec}$ denotes the spectroscopic redshift. We also
restrict the $i$-band magnitude as $i>17$ because bright galaxies
might be saturated in the HSC images. We note that these color cuts are
meant for the rough selection of red-sequence galaxies and does not
need to be strict as any outliers are clipped in the course of the
calibration process \citep[see][]{oguri14}. We also note that these
color cuts are only for spectroscopic galaxies used for the
calibration, i.e., we do not apply any color cuts for photometric
galaxies used for cluster finding.  

\begin{figure}
 \begin{center}
  \includegraphics[width=8cm]{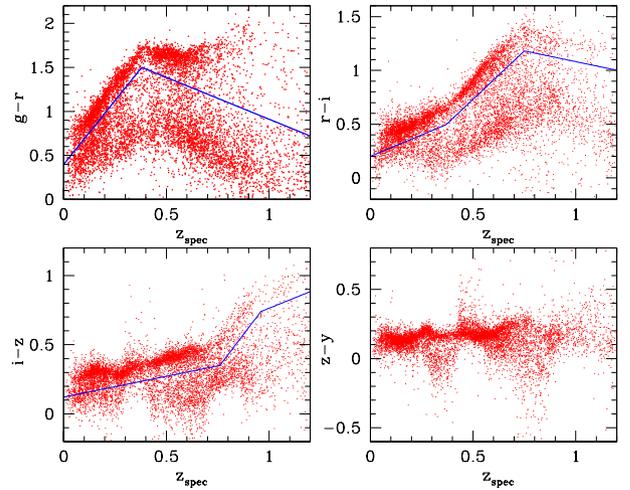} 
 \end{center}
\caption{Colors of spectroscopic galaxies as a function of
  redshift. For illustrative purpose, we plot a subsample of only 5\%
  of all the spectroscopic galaxies. We show $g-r$ ({\it upper left}),
  $r-i$ ({\it upper right}), $i-z$ ({\it lower left}), and $z-y$ ({\it
    lower right}). Solid lines show our color cuts defined by
  equations~(\ref{eq:gr1}), (\ref{eq:ri1}), and
  (\ref{eq:iz1}).}\label{fig:zcol}  
\end{figure}

We show the redshift distribution of the spectroscopic galaxies for
the calibration in Figure~\ref{fig:hist_specz}, and distributions of
colors of the spectroscopic galaxies as well as our color cuts in 
Figure~\ref{fig:zcol}. After the color cuts the number of spectroscopic
galaxies reduces to 62,998. In fact we use only 90\% of the
spectroscopic galaxies after the color cuts for the calibration,
and reserve the remaining 10\% to make sure that we are not grossly
overfitting the data. To check this point, we compute photometric
redshifts of individual spectroscopic galaxies using our stellar
population synthesis (SPS) model with calibrated colors and derive the
bias and scatter of photometric redshifts of these spectroscopic
galaxies. We find that the bias and scatter are similar between
spectroscopic galaxies used for the calibration and those not used for
the calibration. More specifically, we compute the bias and scatter of
the residual $(z_{\rm photo}-z_{\rm spec})/(1+z_{\rm spec})$ for these
spectroscopic galaxies with good SPS model fitting results
($\chi^2<20$ for 4 degree of freedom; this cut is only for checking
the accuracy of photometric redshifts for individual galaxies
presented in this subsection, and in cluster finding we do not apply
this $\chi^2$ cut), and find the bias and scatter of $-0.0017$ and $0.021$ for
galaxies used for the calibration, and $-0.0020$ and $0.021$ for
galaxies not used for the calibration.

\subsection{Updates of the CAMIRA Algorithm}\label{sec:updates}

\begin{figure}
 \begin{center}
  \includegraphics[width=8cm]{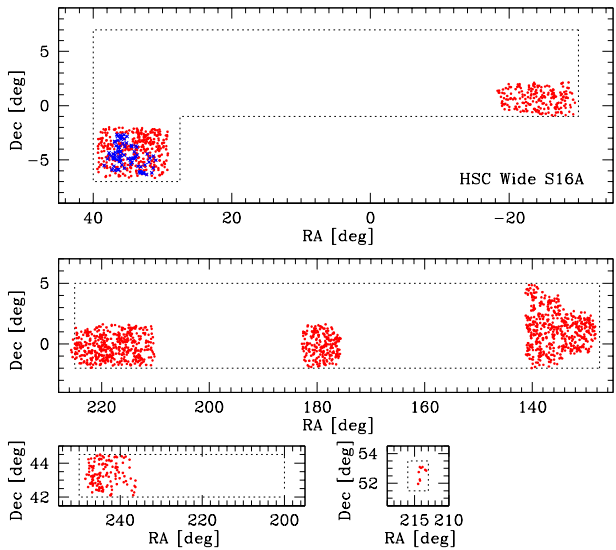} 
 \end{center}
\caption{Spatial distribution of clusters. Filled points show the
  positions of HSC Wide S16A clusters presented in this paper. The
  planned footprint of the entire HSC Wide Survey is the
  region enclosed by dotted lines. We also show the spatial
  distribution of X-ray clusters used in Section~\ref{sec:xray} by
  crosses.
}\label{fig:camira_radec}  
\end{figure}

The detailed methodology of the CAMIRA algorithm was presented in
\citet{oguri14}. In short, it fits all photometric galaxies with
a SPS model of \citet{bruzual03} to compute likelihoods of being
red-sequence galaxies as a function of redshift. The model is
calibrated using spectroscopic galaxies, which are used to derive
residual colors of SPS model fitting as a function of wavelength and
redshift (see also below). Using the likelihoods we define a
non-integer ``number parameter'' for each galaxy as a function of
redshift, such that the sum of number parameters reduces to
richness. Using the number parameter, the three-dimensional richness
map is constructed to locate cluster candidates from peaks in the
richness map.
In CAMIRA, richness describes the number of red member
  galaxies with stellar masses $M_*\gtrsim 10^{10.2}M_\odot$ and
  within a circular aperture with a radius $R\lesssim 1h^{-1}{\rm
    Mpc}$ in physical unit. In fact we use smooth filters for stellar
  masses, $F_M(M_*)\propto
  \exp[-(M_*/10^{13}M_\odot)^4-(10^{10.2}M_\odot/M_*)^4]$, and for spatial
distributions, $F_R(R) \propto
\Gamma[4,(R/R_0)^2]-(R/R_0)^4\exp[-(R/R_0)^2]$ with $R_0=0.8h^{-1}{\rm
Mpc}$, to compute richness. The spatial filter is a compensated filter
such that the background level is automatically subtracted in deriving
richness. 
We then identify a Brightest Cluster Galaxy (BCG) for 
each cluster candidate
by selecting a high stellar mass galaxy near the
  richness peak, and refine richness, cluster photometric
redshift, and the BCG, iteratively. Interested readers are referred to
\citet{oguri14} for more information.  

In this paper, we update the CAMIRA algorithm in several ways and also
adjust model parameters, in order to produce better results in the HSC
Survey which covers wider redshift range than the SDSS. First, while
the central wavelength $\lambda_0$ used in the polynomial fitting of
the residual magnitudes of the SPS model fitting \citep[see equation~3
  in][]{oguri14} was fixed to 5000\AA, here we allow it to vary as a
function of redshift, because the range of the rest frame wavelength
covered by the HSC filters change significantly from $z=0$ to $z\sim
1$. Specifically, we assume the functional form of  
\begin{equation}
  \lambda_0(z)=\lambda_{0,z=0}+\gamma z,
\end{equation}
and adopt $\lambda_{0,z=0}=6400\AA$ and $\gamma=-2850\AA$. Second, we
also include the redshift dependence of the model scatter $\sigma_{\rm
  resi}$. We do so by deriving $\sigma_{\rm resi}$ as a function of
redshift using the spectroscopic galaxies within the redshift bin 
$\Delta z=0.13$. The centering parameter $\sigma_R$ in equation~(19)
of \citet{oguri14} is modified from $\sigma_R=0.3h^{-1}$Mpc
to $0.12h^{-1}$Mpc to reduce the cluster miscentering effect (see
also Section~\ref{sec:miscenter}). Finally, we also update some
parameters related to mask corrections, which takes account of the
effect of small-scale gaps and holes seen in the input galaxy
catalog \citep[see][]{oguri14}. We use new parameter values of
$\theta_{\rm mask}=0\farcm4$ (aperture size for creating a mask map
from the input galaxy catalog), $f_{\rm mask,c}=0.7$, and 
$f_{\rm mask,b}=0.3$ (correction factors for the inner and outer parts
of the spatial filter for cluster finding). 

\begin{figure}
 \begin{center}
  \includegraphics[width=8cm]{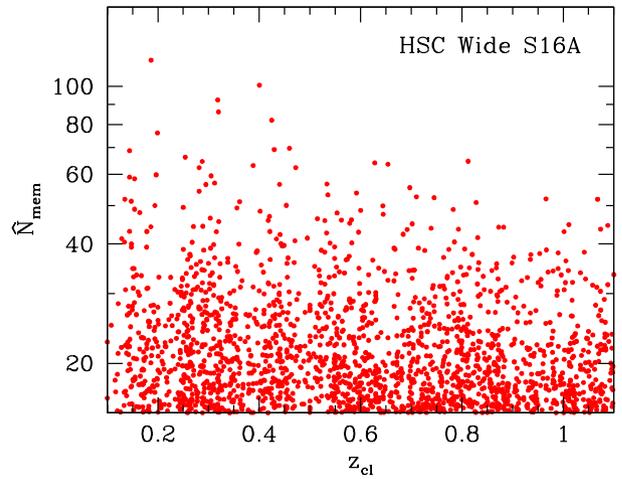} 
 \end{center}
\caption{Distribution of clusters in the richness-redshift plane. The
  cluster catalog is constructed in the redshift range $0.1<z_{\rm
    cl}<1.1$ and the richness range $\hat{N}_{\rm mem}>15$.}\label{fig:camira_nz} 
\end{figure}

\section{Cluster Catalog}\label{sec:catalog}

\subsection{Basic characteristics}\label{sec:basic}

We construct an HSC Wide S16A cluster catalog in the redshift range
$0.1<z_{\rm cl}<1.1$, where $z_{\rm cl}$ denotes photometric
redshifts of clusters. Cluster photometric redshifts are computed in
the course of CAMIRA cluster finding by combining photometric
redshifts of high-confidence cluster member galaxies
\citep[see][]{oguri14}. The upper limit of the redshift comes mainly
from the lack of spectroscopic galaxies for the calibration, but it is
also due to the limited wavelength coverage. The lower limit of the
redshift is because large angular sizes of low-redshift clusters make
cluster finding challenging, and member galaxies also tend to be too
bright in HSC images. We select clusters with the mask-corrected
richness $\hat{N}_{\rm mem}$ \citep[see][for the definition]{oguri14}
higher than $\hat{N}_{\rm mem}=15$. The catalog contains 1921 clusters.

\begin{figure*}
 \begin{center}
  \includegraphics[width=15cm]{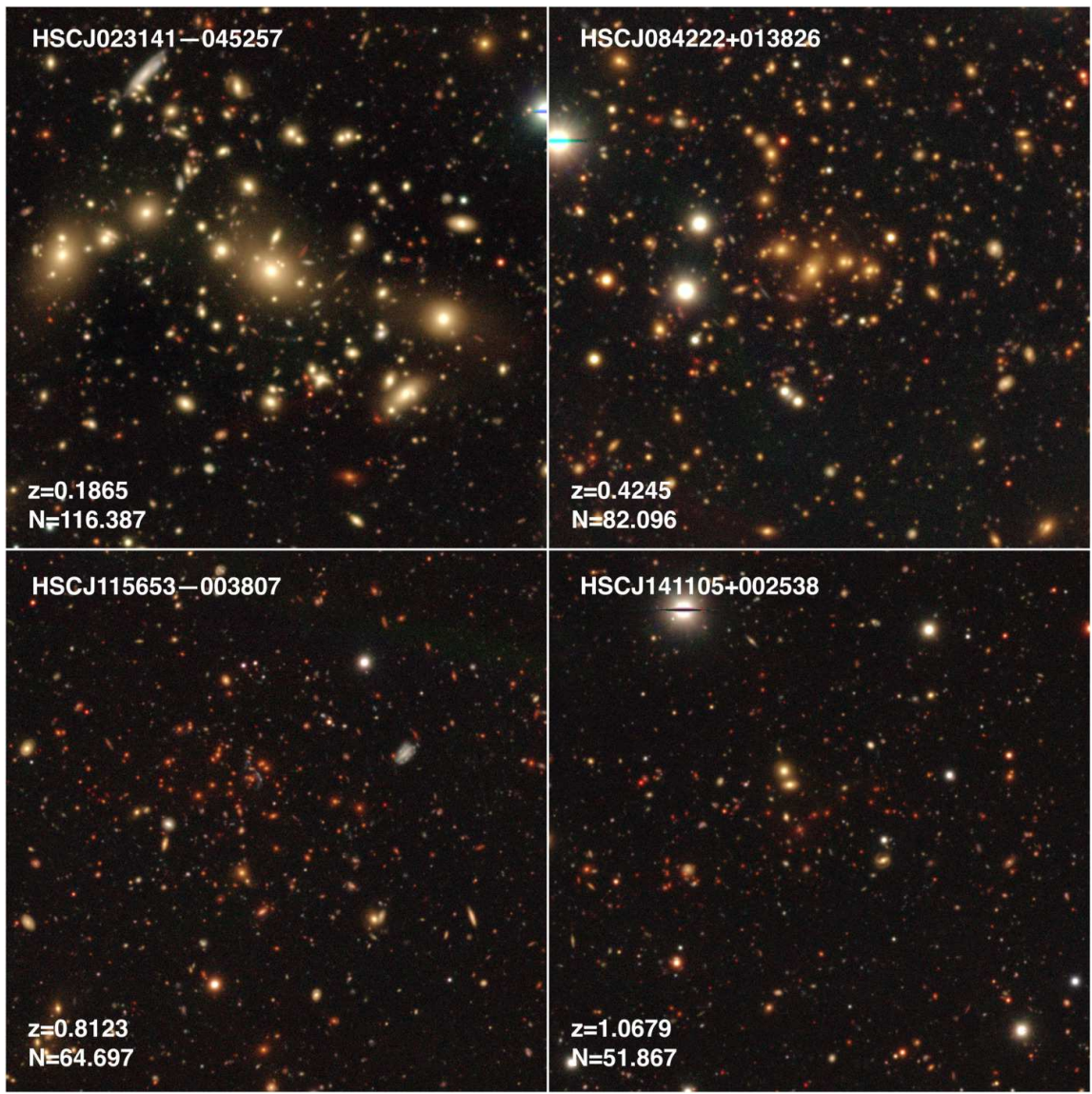} 
 \end{center}
\caption{HSC $grz$-band color composite images of the richest clusters
 at different redshifts in the HSC Wide S16A cluster catalog. The
 cluster photometric redshift $z=z_{\rm cl}$ and richness $N=\hat{N}_{\rm
   mem}$ are indicated in each panel. The size of each panel, which is
  centered at the BCG identified by the CAMIRA algorithm, is
  approximately $3\farcm7 \times 3\farcm7$.
  North is up and east is left.}\label{fig:cl_comb_wcap} 
\end{figure*}

In \citet{oguri14}, the richness $\hat{N}_{\rm mem}$  is further
corrected for the incompleteness of detections of member galaxies. 
This was particularly important for SDSS in which our member
galaxy selection (corresponding to $L\gtrsim 0.2L_*$) was incomplete
at $z\gtrsim 0.3$ due to the shallow depth of SDSS images. In
contrast, the HSC Wide Survey is deep enough to detect almost all
member galaxies used for our cluster finding
(including  the non full-depth area we use in this paper),
which again corresponds to $L\gtrsim 0.2L_*$, even at $z\sim
1.1$. Therefore in this paper we do not apply any richness correction
due to member galaxy incompleteness (i.e., $\hat{N}_{\rm
  cor}=\hat{N}_{\rm mem}$ in the notation of \citealt{oguri14}). We
show the spatial distribution of the HSC Wide S16A clusters in
Figure~\ref{fig:camira_radec}, and the 
distribution in the richness-redshift plane in Figure~\ref{fig:camira_nz}.
The area of the cluster search region of HSC Wide S16A is estimated as
$\sim 232$~deg$^2$. The area is larger than that of the full-color and
full-depth region of 174~deg$^2$ for the HSC Wide S16A dataset because
we use the region that does not reach full depth, as described in
Section~\ref{sec:hscdata}. We show HSC images of the richest clusters
in several different redshift ranges in Figure~\ref{fig:cl_comb_wcap}.

\subsection{Number density of clusters}\label{sec:numden}

\begin{figure}
 \begin{center}
  \includegraphics[width=8cm]{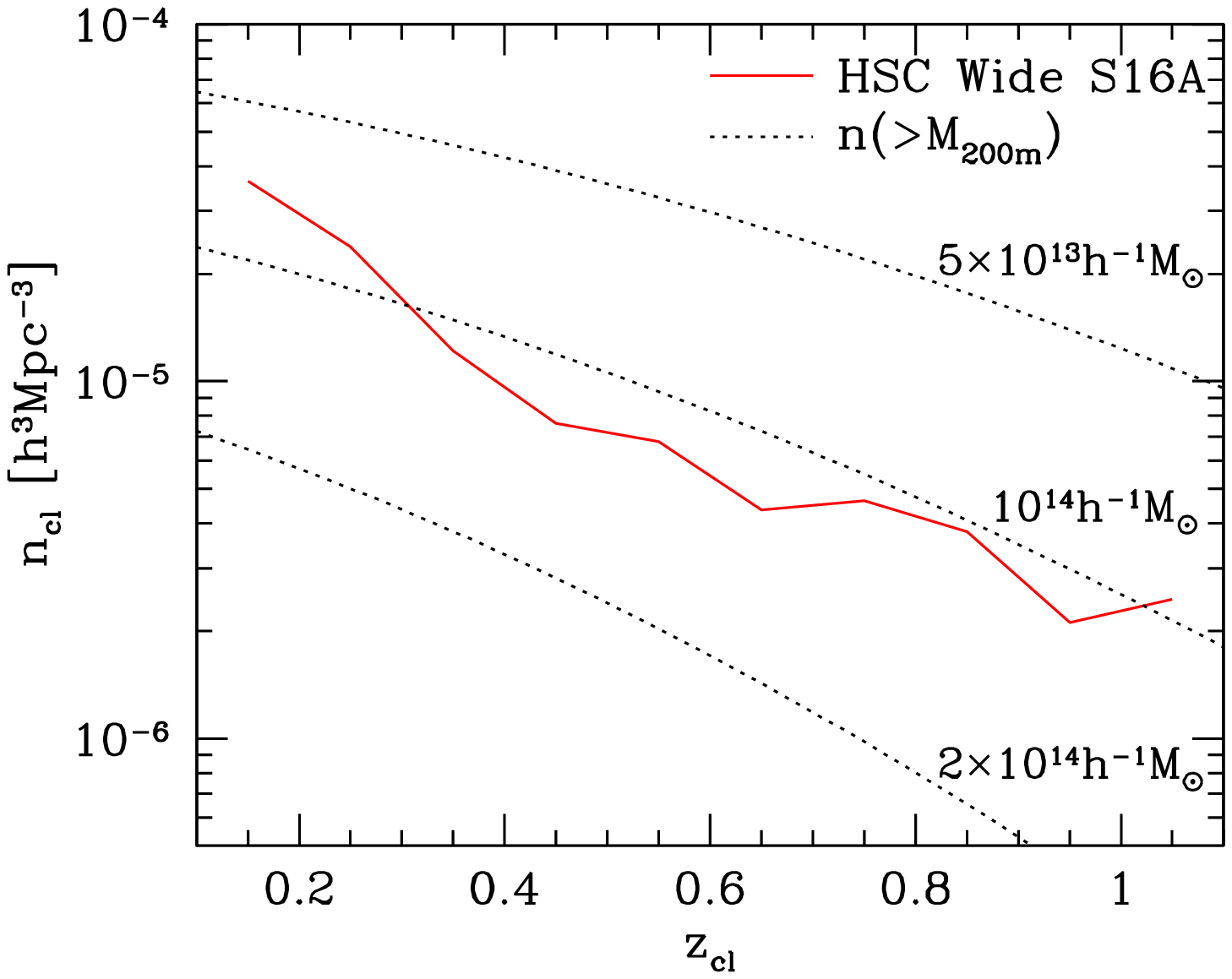} 
 \end{center}
\caption{The comoving number density of HSC Wide S16A clusters as a
  function of cluster redshift ({\it solid}). The Poisson error is
  sufficiently small, $\sim 10$\% in each bin. Dotted lines show
  the predicted number densities of halos with masses $M_{\rm
    200m}>5\times 10^{13}h^{-1}M_\odot$ ({\it top}), 
  $10^{14}h^{-1}M_\odot$ ({\it middle}), and $2\times
  10^{14}h^{-1}M_\odot$ ({\it bottom}), which are computed using a halo
  mass function of \citet{tinker08}.}\label{fig:camira_ncl} 
\end{figure}

One way to infer the mass of a cluster catalog is to
  compare its number density with a number density of halos for a
  given cosmological model \citep[e.g.,][]{jimeno17}.
We show the comoving number density of HSC Wide S16A clusters as a
function of redshift in Figure~\ref{fig:camira_ncl}. The number
density declines slowly with increasing redshift, which is consistent
with the DES Science Verification (SV) data redMaPPer cluster catalog
\citep{rykoff16} and implies the redshift evolution of the mass
threshold is not strong. In order to check this point further, we
compute the predicted number densities of halos with constant mass
threshold using the halo mass function of \citet{tinker08}. Here we
adopt the mean overdensity mass $M_{\rm 200m}$ which is defined as the
mass enclosed within a sphere of radius $r_{\rm 200m}$ within which
the mean density is 200 times the mean matter density of the
Universe. The comparison shown in Figure~\ref{fig:camira_ncl} suggests 
that our richness threshold of $\hat{N}_{\rm mem}>15$ roughly
corresponds to the mass threshold of 
$M_{\rm 200m}\gtrsim 10^{14}h^{-1}M_\odot$. 

\subsection{Performance of photometric redshifts}

\begin{figure}
 \begin{center}
  \includegraphics[width=8cm]{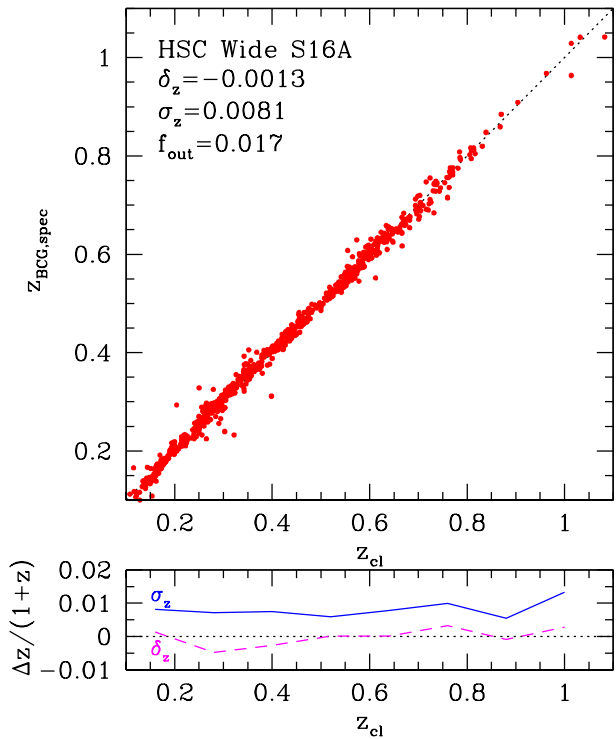} 
 \end{center}
\caption{{\it Upper:} The comparison between cluster photometric
redshifts $z_{\rm cl}$ and spectroscopic redshifts of BCGs 
 $z_{\rm BCG,spec}$. {\it Lower:} The bias $\delta_z$ ({\it dashed}) 
 and scatter $\sigma_z$ ({\it solid}) 
 of the residual $(z_{\rm cl}-z_{\rm BCG,spec})/(1+z_{\rm BCG,spec})$
 as a function of redshift. 
}\label{fig:comp_clspec} 
\end{figure}

Photometric redshifts of clusters, $z_{\rm cl}$, are derived on the
course of cluster finding \citep[see][]{oguri14}. In order to check
the accuracy of cluster photometric redshifts, we cross-match the HSC
Wide S16A cluster catalog with the spectroscopic galaxy catalog
(Section~\ref{sec:specgal}). We find that BCGs of 843 clusters have
spectroscopic redshifts, $z_{\rm BCG,spec}$. 
Figure~\ref{fig:comp_clspec} compares $z_{\rm cl}$ and 
$z_{\rm BCG,spec}$, which clearly shows that cluster photometric
redshifts are quite accurate for the whole redshift range. To quantify
the accuracy of cluster photometric redshifts, we compute the residual 
$(z_{\rm cl}-z_{\rm BCG,spec})/(1+z_{\rm BCG,spec})$ for all the
clusters and define the bias $\delta_z$ and scatter $\sigma_z$ by 
average and standard deviation of the residual with $4\sigma$
clipping. We then define the outlier rate $f_{\rm out}$ as the
fraction of galaxies that are removed by the $4\sigma$ clipping. 
Using all the galaxies, we find the bias $\delta_z=-0.0013$, the
scatter $\sigma_z=0.0081$, and the outlier rate $f_{\rm out}=0.017$.
In Figure~\ref{fig:comp_clspec} we also show the bias $\delta_z$ and
scatter $\sigma_z$ as a function of redshift, finding that the bias is
$|\delta_z|<0.005$ and the scatter is $\sigma_z<0.01$ for most of the
redshift range. This performance of cluster photometric redshifts
is comparable to that of SDSS redMaPPer and CAMIRA clusters
\citep{rykoff14,oguri14} and better than that of DES SV redMaPPer
clusters \citep{rykoff16}. We note that the availability of a large
sample of spectroscopic galaxies from SDSS and other spectroscopic
surveys (see Section~\ref{sec:specgal}) is an advantage of the HSC
Survey over DES. However, as is clear from Figure~\ref{fig:comp_clspec}, 
there are not many spectroscopic galaxies at $z\gtrsim 0.8$, which
suggests the importance of spectroscopic follow-up observations of
those high-redshift clusters to test their reliability further.

\section{Comparison with SDSS}\label{sec:sdss}

\begin{figure}
 \begin{center}
  \includegraphics[width=8cm]{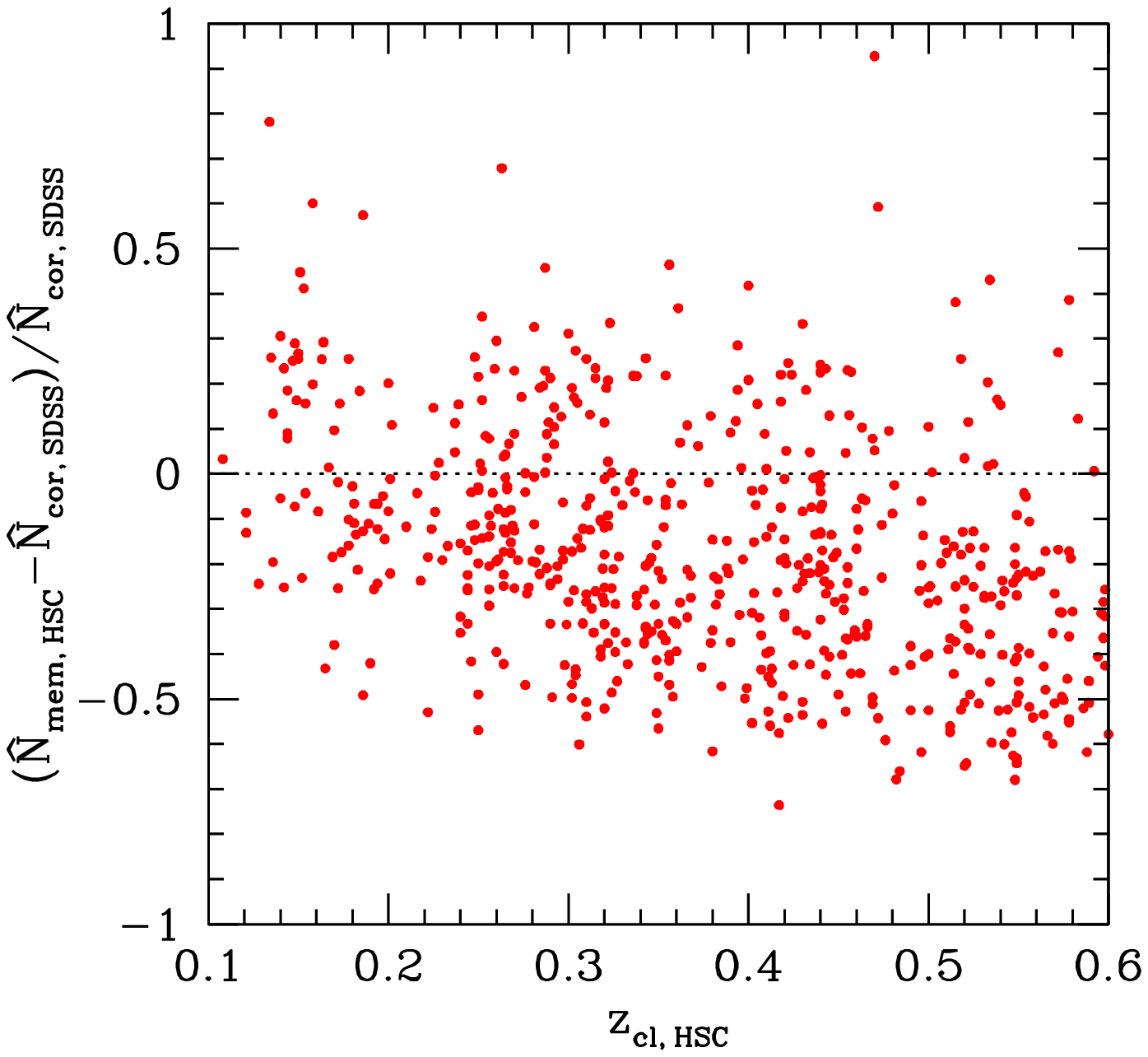} 
 \end{center}
\caption{Comparison of the CAMIRA richness in HSC, $\hat{N}_{\rm mem,HSC}$,
  with the incompleteness-corrected CAMIRA richness in SDSS,
  $\hat{N}_{\rm cor, SDSS}$. The fractional
  difference of these two richness is
  plotted as a function of cluster redshift from HSC, $z_{\rm cl, HSC}$.
}\label{fig:check_dr8_dnz} 
\end{figure}

Since the HSC data is much deeper than SDSS, it is expected that we
can identify cluster member galaxies more reliably in HSC, down to
the stellar mass limit for the richness calculation, $\sim
10^{10.2}M_*$, which roughly corresponds to the luminosity range of
$L\gtrsim 0.2L_*$ recommended in \citet{rykoff12}.
Due to the shallowness of the SDSS data, in \citet{oguri14} we applied
a richness correction factor $f_N(z)=\hat{N}_{\rm mem}/\hat{N}_{\rm
  cor}$ to account for the incompleteness of member galaxy
detections. In contrast, as discussed in Section~\ref{sec:basic}, in
HSC we do not apply any correction for the member galaxy
incompleteness simply because HSC data are deep enough to detect all
the member galaxies of interest out to $z\sim 1.1$. Thus we compare
the richness in SDSS and HSC to check the accuracy of the
incompleteness correction. 

We cross match the CAMIRA SDSS catalog with the HSC Wide S16A
catalog. Our matching criterion is that clusters whose physical
transverse distance is within 1$h^{-1}{\rm Mpc}$ and the difference in
cluster photometric redshifts is smaller than $0.1$. For the SDSS
cluster catalog, we use an updated (v1.2) CAMIRA SDSS cluster catalog
which slightly differs from the cluster catalog published in
\citet{oguri14}. The updated CAMIRA SDSS catalog\footnote{The updated
  CAMIRA SDSS cluster catalog (version 1.2) is available at
  http://www.slac.stanford.edu/\~{}oguri/cluster/.} adopts the new 
centering parameters described in Section~\ref{sec:updates}, and
contains 83735 clusters in the redshift range $0.1<z_{\rm cl}<0.6$ and
with the corrected richness $\hat{N}_{\rm cor}>20$.

In Figure~\ref{fig:check_dr8_dnz}, we compare the richness from HSC
with the incompleteness-corrected richness from SDSS as a function of
cluster redshift. While at low redshift these richness agree with each
other, at higher redshift ($z\gtrsim 0.3$), where the richness
correction is applied to SDSS richness, we find a systematic offset
between these two richness, such that richness in HSC is
systematically lower than richness in SDSS. This comparison suggests
a possible systematic bias in the incompleteness correction in the
CAMIRA SDSS cluster catalog, and hence highlights importance of using
deep data to detect sufficiently faint cluster members for optical
cluster finding. The deep photometric data are also important to study
the redshift evolution of the luminosity function of cluster member
galaxies. 

\section{Comparison with X-ray cluster catalogs}\label{sec:xray} 

\subsection{X-ray catalogs}\label{sec:xxl}

The footprint of the HSC Wide S16A cluster catalog has a large overlap
with two large area X-ray surveys, the XMM Large Scale Structure
survey \citep[XMM-LSS;][]{pierre04} and the XXL survey \citep{pierre16}.
We adopt a sample of 52 X-ray bright clusters selected from the
11~deg$^2$ XMM-LSS survey region presented in \citet{clerc14}, and
also a sample of 51 X-ray bright clusters from the $\sim 25$~deg$^2$
XXL North survey region presented in \citet{pacaud16} and \citet{giles16}.
For both the cluster catalogs, we use X-ray centroids, redshifts
from the spectroscopy of member galaxies, X-ray temperatures
$T_X$, and rest-frame [0.5--2]~keV luminosities within 
$r_{500,{\rm MT}}$, $L_{X,500}$, where $r_{500,{\rm MT}}$ is the
radius within which the mean matter density of the cluster becomes 500
times the critical density of the Universe at the cluster redshift
with the corresponding mass $M_{500}$ estimated from a
mass-temperature relation. In the XXL survey, X-ray temperatures
$T_X$ are computed within a fixed aperture of 300~kpc. These two
cluster catalogs have some overlap, and in this paper we use values
from the XXL catalog when a cluster is included on both the XXL and
XMM-LSS catalogs, because the XXL survey is an extension of the
XMM-LSS survey. In addition, we remove 4 XXL cluster located at ${\rm
  Dec}<-6.6$ as the HSC Wide S16A cluster catalog does not cover 
that area (see Figure~\ref{fig:camira_radec}). 

By combining these two X-ray cluster catalogs, we construct a sample of
77 X-ray bright clusters which are used for characterizing the HSC
Wide S16A cluster catalog. The spatial distribution of these X-ray
clusters is shown in Figure~\ref{fig:camira_radec}. 

\subsection{Correlation of richness with X-ray properties}

\begin{figure}
 \begin{center}
  \includegraphics[width=8cm]{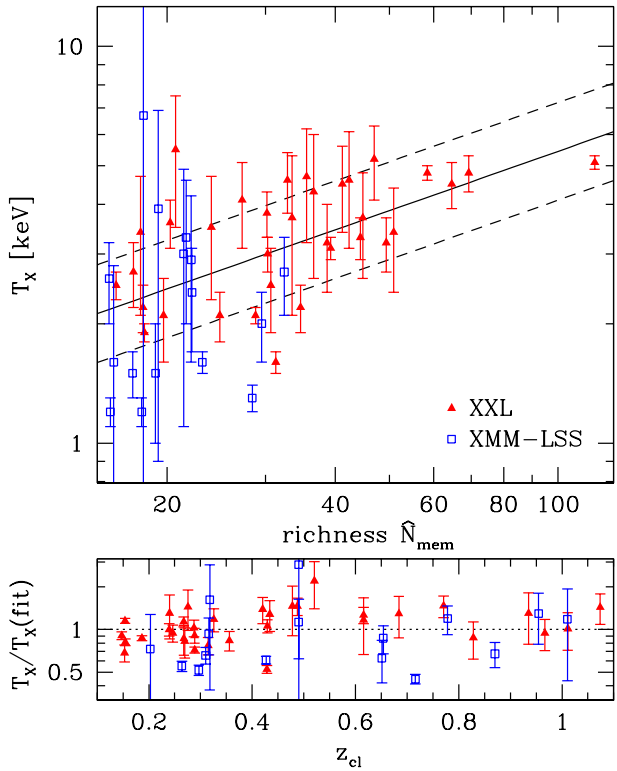} 
 \end{center}
\caption{Comparison between richness $\hat{N}_{\rm mem}$ and X-ray
  temperature $T_X$ for XXL ({\it filled triangles}) and XMM-LSS ({\it
    open squares}) clusters. The solid and dashed lines show the
  best-fit richness-temperature relation and the range of $\pm
  1\sigma$ intrinsic scatter. The lower panel shows the residual of
  fitting as a function of redshift.
}\label{fig:check_tx} 
\end{figure}

\begin{figure}
 \begin{center}
  \includegraphics[width=8cm]{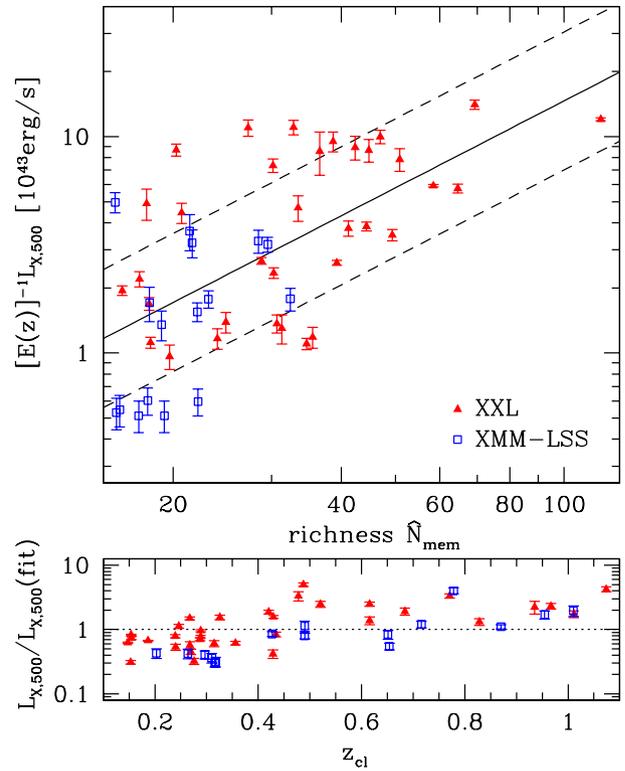} 
 \end{center}
\caption{Same as Figure~\ref{fig:check_tx}, except that the comparison
  between richness $\hat{N}_{\rm mem}$ and evolution-corrected X-ray
  luminosity $\left[E(z)\right]^{-1}L_{X,500}$ is shown. 
}\label{fig:check_lx}  
\end{figure}

The comparison between the richness and X-ray properties such as
the luminosity $L_{X,500}$ and temperature $T_X$ is useful. Because
X-ray properties are more tightly correlated with cluster masses
\citep[but see also][]{andreon10,andreon15}, the
correlation of richness with X-ray properties is often used as a proxy
for the tightness of the correlation between the richness and the
cluster mass \citep[e.g.,][]{rykoff12,rozo14a,oguri14,wen15}.  

We cross match the HSC Wide S16A cluster catalog with the X-ray
cluster catalog constructed in Section~\ref{sec:xxl}. Again, we use a simple
matching criterion that clusters whose physical transverse distance is
within 1$h^{-1}{\rm Mpc}$ and redshift difference is smaller than
$0.1$ are matched. When there are several matching candidates, we
adopt clusters with the smallest transverse separation. Among the 77
X-ray bright clusters from XXL and XMM-LSS, 50 clusters are found to
have counterparts in the HSC Wide S16A catalog. Some of the X-ray
clusters have no counterpart in the HSC cluster catalog simply
because their redshifts fall outside the redshift range of the HSC
Wide S16A catalog (i.e., $z<0.1$ or $z>1.1$). There are other X-ray
clusters that fall within the redshift range of the CAMIRA catalog but
have no counterpart, which are used to estimate the completeness of
the HSC cluster catalog  (see Section~\ref{sec:compl}). 

Figure~\ref{fig:check_tx} compares the richness $\hat{N}_{\rm mem}$
with X-ray temperature $T_X$ for the 50 clusters. There is a clear
positive correlation between $\hat{N}_{\rm mem}$ and $T_X$. To
quantify the correlation, we fit the relation assuming the following
power law model:
\begin{equation}
\log\left(\frac{T_X}{{\rm keV}}\right)=a_T\log\left(\frac{\hat{N}_{\rm
    mem}}{30}\right)+b_T.
\end{equation}
This model is same as that used in \citet{oguri14} except that the
pivot richness is changed to 30. We find the best-fit parameters of
$a_T=0.50\pm0.12$ and $b_T=0.48\pm0.02$. The slope $a_T=0.50$ is
shallower than $2/3$ predicted from $M \propto \hat{N}_{\rm mem}$ and
the self-similar model $T_X\propto M^{2/3}$. The relation also implies
that the mass limit of the HSC cluster catalog of $M_{500c}\gtrsim 2
\times 10^{14}h^{-1}M_\odot$ \citep{lieu16} which is broadly
consistent with the mass limit from the number density
(Section~\ref{sec:numden}). We also find $1\sigma$ intrinsic scatter
of $0.12$ in $\log T_X$, which is comparable to the scatter for the
SDSS CAMIRA and redMaPPer cluster catalogs
\citep{rozo14a,oguri14}. Figure~\ref{fig:check_tx} also suggests that
there is no strong redshift evolution of this relation. 

We then compare $\hat{N}_{\rm mem}$ with X-ray luminosity $L_{X,500}$
for the same clusters in Figure~\ref{fig:check_lx}. It has been shown
that the $L_X$-$T_X$ relation evolves with redshift such that
$E(z)^{-\gamma_{LT}} L_X \propto T_X^B$, where
$E(z)=\sqrt{\Omega_M(1+z)^3+\Omega_\Lambda}$, $B\sim 3$, and the
self-similar model predicts $\gamma_{LT}=1$ \citep[e.g.,][]{giles16}.
We include the effect of redshift evolution by studying the
correlation between $\hat{N}_{\rm mem}$ and evolution-corrected X-ray 
luminosity $\left[E(z)\right]^{-1}L_{X,500}$ assuming the
self-similar evolution $\gamma_{LT}=1$. As before,
Figure~\ref{fig:check_lx} indicates a good correlation between the
richness and X-ray luminosity. We quantify this by fitting it to a
power-law model: 
\begin{equation}
\log\left(\frac{\left[E(z)\right]^{-1}L_{X,500}}{10^{43}{\rm
    erg/s}}\right)=a_L\log\left(\frac{\hat{N}_{\rm mem}}{30}\right)+b_L,
\end{equation}
and find $a_L=1.33\pm0.24$, $b_L=0.47\pm0.05$, and the intrinsic
scatter $\sigma=0.32$.

The residual of fitting as a function of redshift shown in
Figure~\ref{fig:check_lx} indicates a clear trend with redshift. This
may suggest a redshift evolution stronger than the self-similar
prediction, i.e., $\gamma_{LT}>1$ \citep[see also][]{giles16}, although it
must be partly due to the Malmquist bias. This is because the X-ray
cluster sample is a flux-limited sample and X-ray luminosities are
more directly related to X-ray fluxes. The idea is supported by the
fact that XMM-LSS clusters, which have the lower X-ray flux limit,
tend to have lower luminosities than those of XXL clusters at all
redshift range. Hence in order to derive the intrinsic scaling
relation we need to take account of flux limits of these X-ray
surveys, which we leave for future work.

\subsection{Completeness estimation}\label{sec:compl}

\begin{figure}
 \begin{center}
  \includegraphics[width=8cm]{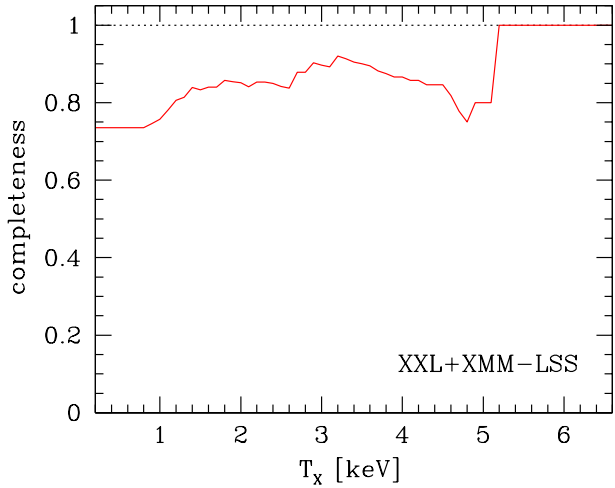} 
 \end{center}
\caption{Completeness of HSC Wide S16A clusters which is estimated
  using X-ray clusters. The completeness is defined by the fraction of
  X-ray clusters that match the HSC cluster catalog to all X-ray
  clusters. The completeness is estimated as a function of X-ray
  temperature threshold (i.e., for each $T_X$ all X-ray clusters with
  $>T_X$ are used to derive the completeness). 
}\label{fig:compl} 
\end{figure}

In addition to using mocks (see Section~\ref{sec:mock}), we estimate the
completeness of the HSC Wide S16A catalog using the 
X-ray cluster catalog. Here we adopt a simple approach to estimate the
completeness by the fraction of X-ray clusters with optical
counterparts to all X-ray clusters. Given an ambiguity of matching
between HSC and X-ray clusters, the completeness presented here should
be interpreted with caution. We derive the completeness as a function
of X-ray temperature threshold; for a given $T_X$, we use all X-ray
cluster above $T_X$ to derive the completeness. We also restrict the
redshift range of the X-ray clusters to $0.1<z<1.1$ for the completeness
estimation.

Figure~\ref{fig:compl} plots the estimated completeness as a function
of X-ray temperature threshold. We find that the completeness is
generally high, $\sim 0.8-0.9$. There are a couple of X-ray clusters
with relatively high X-ray temperatures, $T_X\sim 5$~keV, which do not
have counterparts in the HSC cluster catalog. We find that these X-ray
clusters have companion X-ray clusters separated by $\sim 8'$ at
similar redshifts. For example, one of them corresponds to a member
of a supercluster of galaxies at $z=0.43$ \citep{pompei16}. Since the
CAMIRA algorithm adopts a compensated spatial filter for creating a
richness map \citep{oguri14}, detections of companion clusters near
massive clusters tend to be suppressed, as also noted in
\citet{miyazaki15}. The current analysis of completeness is limited by
a small number of X-ray sample, and an extended analysis using a
larger sample of X-ray clusters is useful for quantifying the impact
of paired clusters on the completeness.  
  
\subsection{Positional offset distribution}\label{sec:miscenter}

\begin{figure}
 \begin{center}
  \includegraphics[width=8cm]{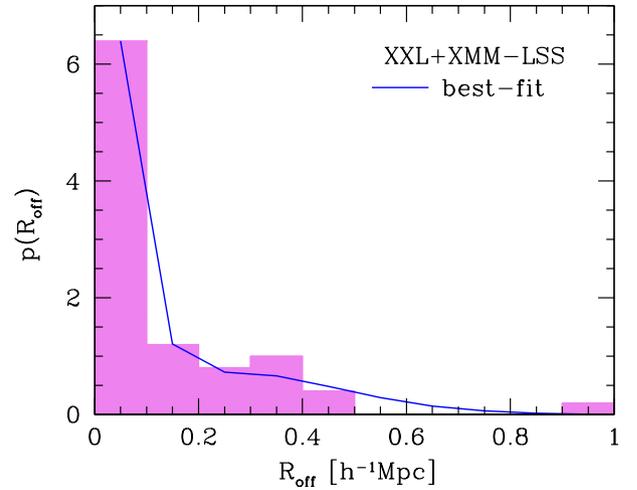} 
 \end{center}
\caption{Distribution of positional offset between HSC Wide S16A
  clusters and X-ray clusters, which is derived using 50 X-ray
  clusters from XXL and XMM-LSS surveys that matches the HSC cluster
  catalog. The offsets $R_{\rm off}$ are defined by physical
  transverse distances between HSC and X-ray cluster centers.
  The solid line shows the best-fit model assuming the functional form 
  of equation~(\ref{eq:poff}).
}\label{fig:offdist} 
\end{figure}

It has been known that optical cluster finders often misidentify
centers of clusters. This miscentering effect is important for
applications of optically-selected clusters such as weak lensing analysis.
Miscentering of optically-selected clusters have indeed been studied with various
approaches, including offsets between optical and X-ray cluster
centers \citep[e.g.,][]{lin04,mahdavi13,rozo14a,oguri14,rykoff16}, offsets
between optical and SZ cluster centers
\citep[e.g.,][]{song12,sehgal13,sifon13,saro15}, and weak lensing
\citep[e.g.,][]{oguri10,george12,oguri14,ford14,viola15,vanuitert16,miyatake16}.
In this paper we use offsets between HSC Wide S16A clusters, whose
centers are defined by the locations of BCGs identified by the CAMIRA
algorithm \citep[see][]{oguri14}, and the X-ray clusters from XXL and
XMM-LSS to study the miscentering effect.  

Figure~\ref{fig:offdist} shows the offset (physical transverse
distance $R_{\rm off}$) distribution derived from the 50 X-ray
clusters that are matched with the HSC Wide S16A cluster catalog. 
In the HSC Wide S16A cluster catalog, cluster centers are defined by
centroids of BCGs identified by the CAMIRA algorithm.
The distribution is clearly peaked at $R_{\rm off}\sim 0$, indicating that
HSC Wide S16A clusters are well centered on average. On the other
hand, there is also a tail toward large $R_{\rm off}$, as has been
found in other optically-selected cluster catalogs. We fit the measured
distribution to the following two-component Gaussian model
\citep[e.g.,][]{oguri11}
\begin{eqnarray}
p(R_{\rm off})&=&f_{\rm cen}\frac{R_{\rm
    off}}{\sigma_1^2}\exp\left(-\frac{R_{\rm off}^2}{2\sigma_1^2}\right)
\nonumber\\
&&+(1-f_{\rm cen})\frac{R_{\rm
    off}}{\sigma_2^2}\exp\left(-\frac{R_{\rm off}^2}{2\sigma_2^2}\right).
\label{eq:poff}
\end{eqnarray}
We simultaneously fit the fraction of well-centered clusters,
$f_{\rm cen}$, the standard deviation for the well-centered
population, $\sigma_1$, and the standard deviation for the miscentered
population, $\sigma_2$. The best-fit parameters are $f_{\rm cen}=0.68\pm0.09$,
$\sigma_1=0.046\pm0.009 h^{-1}{\rm Mpc}$, and
$\sigma_2=0.26\pm0.04 h^{-1}{\rm Mpc}$.
We confirmed that fitting results with smaller bin
  sizes are consistent with the result above within the error bars.
The best-fit model shown in Figure~\ref{fig:offdist} indicates that
the two-component model fits the observation reasonably well.

We check the possible richness dependence of the offset distribution
by dividing the cluster catalog into two richness bins, $\hat{N}_{\rm
  mem}<30$ and  $\hat{N}_{\rm mem}>30$, and repeating the analysis
above. Given the smaller number of X-ray clusters in each richness
bin, we fix $\sigma_1$ and $\sigma_2$ to the best-fit values obtained
from the analysis of the full sample, and fit only $f_{\rm cen}$. We
find that $f_{\rm cen}=0.66\pm 0.10$ for $\hat{N}_{\rm mem}<30$, and
$f_{\rm cen}=0.71\pm 0.12$ for $\hat{N}_{\rm mem}>30$, which indicates
that there is no strong dependence of the offset distribution on
richness. 

\section{Analysis of mock galaxy samples}\label{sec:mock}

\subsection{Construction of mock galaxy samples}\label{sec:genmock}

We create mock galaxy samples specifically designed for testing cluster
finding algorithms. Since accurate galaxy colors are crucial for
this purpose, we use colors of true observed galaxies in the HSC
Survey. We use the HSC galaxy catalog in the COSMOS \citep{scoville07}
field for the field galaxy population, and use CAMIRA clusters in HSC
Wide S16A to select cluster member galaxies. $N$-body simulations 
are used to produce realistic distributions of halo masses and
redshifts. Thus our mock galaxy catalogs use realistic cluster
distributions, and also contain gaps and holes in the spatial
distribution of galaxies due to bright stars and bad columns, which
allow realistic characterization of completeness and purity. We
provide a detailed description of our mock galaxy samples in Appendix~3.

We note that we need to make assumptions on the input richness-mass
relation, red galaxy fraction, and their redshift evolution in order
to create these mock galaxy samples, which may not necessarily be
correct. While quantitative results such as the scatter in the
observed richness-mass relation and the completeness as a function of
halo masses depend on these assumptions, we expect that the
qualitative behavior of the completeness and purity estimated from
these mocks is not very sensitive to these assumptions.  

\subsection{Analysis results}

\begin{figure}
 \begin{center}
  \includegraphics[width=8cm]{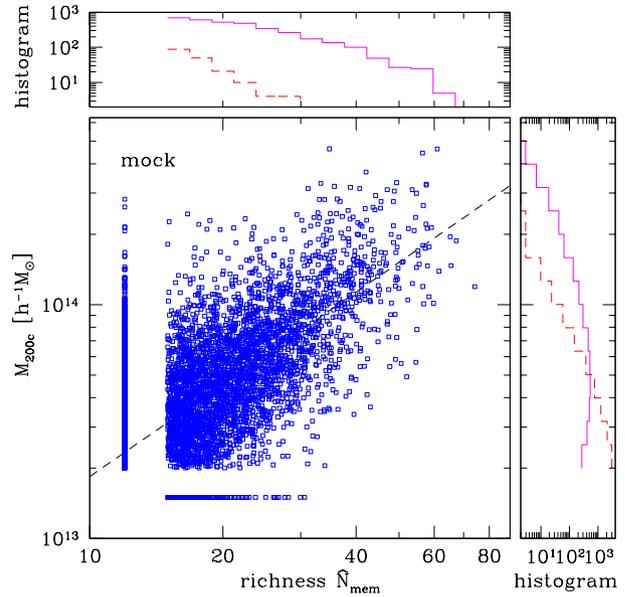} 
 \end{center}
\caption{Comparison of cluster richness $\hat{N}_{\rm mem}$ and
  halo mass $M_{\rm 200c}$ from the analysis of mock galaxy catalogs
  (see Section~\ref{sec:genmock} for the description). For 
  illustrative purpose, we assign $\hat{N}_{\rm mem}=12$ for halos
  without any counterpart in the cluster catalog, and
  $M_{\rm 200c}=1.5\times 10^{13}h^{-1}M_\odot$ for clusters without any
  counterpart in the halo catalog. The dashed line
  shows a power-law fit to $\langle \log M|N\rangle$ using the matching
  result with $\hat{N}_{\rm mem}>20$ (see
  equation~\ref{eq:nmpow_mock}). In the upper and right panels, we
  also show the histograms of matched ({\it solid}) and unmatched 
  ({\it dashed}) clusters/halos.
}\label{fig:mock_nm} 
\end{figure}

We run the CAMIRA algorithm on the mock galaxy samples using the same
setup as the one used to create the HSC Wide S16A catalog. We then
cross match CAMIRA cluster catalogs on the mocks with input
halo catalogs. We use the matching criterion of the physical
transverse distance being smaller than 1$h^{-1}{\rm Mpc}$ and the
difference between the halo redshift and the cluster photometric
redshift being smaller than $0.1$. Given the limited range of the
redshift and halo mass of our mock galaxy samples (see Appendix~3), we
also restrict the redshift range of halos to $0.3<z_{\rm halo}<1.0$
and the mass range to $M_{\rm 200c}>2\times 10^{13}h^{-1}M_\odot$ in
cross-matching. We record all clusters and halos that fall within the
same redshift range but fail to match, which are used to estimate
completeness and purity below. We analyze 90~realizations of the
COSMOS-size mock galaxy samples, leading to the total area
of $\sim 144$~deg$^2$.

Figure~\ref{fig:mock_nm} shows the result of matching. As expected we
can clearly see a positive correlation between halo masses and
richness. We find that most of the CAMIRA clusters in the mocks have
counterparts in the halo catalog except for the very low richness end,
which indicates that the purity of the CAMIRA cluster catalog, which
is defined by the fraction of CAMIRA clusters that corresponds to real
massive halos, is high, $>0.95$ down to the richness limit of
$\hat{N}_{\rm mem}=15$. On the other hand, Figure~\ref{fig:mock_nm}
indicates that there are halos up to $\sim 10^{14}h^{-1}M_\odot$ that
are not detected by the CAMIRA algorithm, which implies that the
completeness, which is defined by the fraction of halos that are
correctly identified by the CAMIRA algorithm, requires careful
studies.   
   
Our CAMIRA cluster sample is constructed in the limited richness range
of $\hat{N}_{\rm mem}>15$, which may impact on the estimate of the
completeness as a function of halo mass. We derive a simple correction
of the effect of the lower limit of the richness by assuming a
log-normal form for the mass-richness relation 
\begin{equation}
p(\log M|N)=\frac{1}{\sqrt{2\pi}\sigma_{\log
    M}}\exp\left[-\frac{(\log M-\langle\log M|N\rangle)^2}{2\sigma_{\log M}^2}\right],
\end{equation}
here we adopt simplified notations, $N=\hat{N}_{\rm mem}$ and
$M=M_{200c}$ in units of $h^{-1}M_\odot$. For the median richness-mass
relation $\langle\log M|N\rangle$, we assume the following power-law
form
\begin{equation}
  \langle\log M|N\rangle =a_M\log\left(\frac{N}{30}\right)+b_M.
  \label{eq:nmpow_mock}
\end{equation}
To derive the parameters, we use the matching result only with $N>20$
in order to reduce the effect of lower mass limit and perform fitting.
The resulting parameters are $a_M=1.31$, $b_M=13.89$, and
$\sigma_{\log M}=0.19$. We also confirm that the log-normal model
reproduces the observed distribution of halo masses around the median
richness-mass relation reasonably well. 

The Bayes theorem suggests that $p(\log M|N)$ can be converted to
$p(\log N|M)$ via
\begin{equation}
p(\log N|M)=\frac{p(\log M|N)p(\log N)}{\int d(\log N)p(\log M|N)p(\log N)}.
\end{equation}
Using the locally power-law model \citep{rozo14b}, it is shown that
this distribution has a log-normal distribution with the mean and
variance of
\begin{equation}
  \langle\log N|M\rangle=\frac{\log M -b_M}{a_M}
  -\frac{\beta\ln 10}{a_M^2}\sigma_{\log M}^2,
\end{equation}
\begin{equation}
\sigma_{\log N}=\frac{1}{a_M}\sigma_{\log M},
\end{equation}
where $\beta$ is the slope of the richness function, $p(\log N)
\propto dn/d\log N \propto N^{-\beta}$. We assume $\beta=1$ for
simplicity, which is broadly consistent with the observed richness
distribution near the richness threshold. We can then compute the
selection function (or completeness) for a sample with $N>N_{\rm
  lim}$ predicted by the log-normal model as 
\begin{eqnarray}
S(M| N>N_{\rm lim})&=&\frac{p(\log M|N>N_{\rm lim})}{p(\log
  M)}\nonumber\\
&=&\int_{N_{\rm lim}}^\infty p(\log N|M) d(\log N)\nonumber \\
&=& \frac{1}{2}{\rm erfc}\left[\frac{\log
    N_{\rm lim}-\langle\log N|M\rangle}{\sqrt{2}\sigma_{\log
      N}}\right].
\end{eqnarray}
The completeness we are interested in here is the completeness that
cannot be explained by a simple extrapolation of the mass-richness
relation. To derive this, we first prepare a sample of CAMIRA clusters
in the mocks with $N>N_{\rm lim}$ that successfully match to the halo
catalog. We denote the number distribution of these clusters as
$n_{\rm obs}(M|N>N_{\rm lim})$. From the original halo catalog, we can
also derive the number distribution of all halos, $n(M)$. We estimate
the completeness, for which the effect of the lower limit of richness is
corrected, as
\begin{equation}
{\rm completeness}=\frac{n_{\rm obs}(M|N>N_{\rm lim})}{n(M)S(M|
  N>N_{\rm lim})}.
\label{eq:compl}
\end{equation}
Put another way, this provides the incompleteness of identifying halos
that cannot be explained by a simple extrapolation of the
mass-richness relation. 

\begin{figure}
 \begin{center}
  \includegraphics[width=8cm]{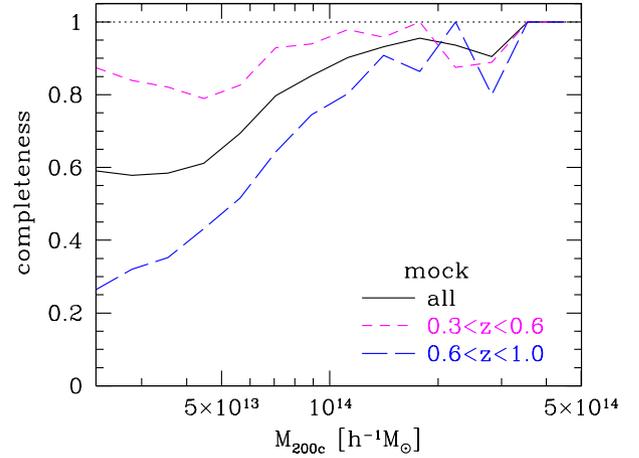} 
 \end{center}
\caption{Completeness estimated from the analysis of mock galaxy
  catalogs as a  function of halo mass $M_{\rm 200c}$. The
  completeness is defined by equation~(\ref{eq:compl}). The solid line
  shows the completeness using all halos. Short-dashed and long-dashed
  lines indicate the completeness for subsamples of halos at
  $0.3<z<0.6$ and $0.6<z<1.0$, respectively.
}\label{fig:mock_compl} 
\end{figure}

Figure~\ref{fig:mock_compl} shows the completeness defined by
equation~(\ref{eq:compl}). Here we adopt $N_{\rm lim}=15$, although the
result is insensitive to the specific choice of $N_{\rm lim}$ as
expected. We note that, for $N_{\rm lim}=15$, the correction factor
$S(M)$ is very close to unity down to 
$M_{\rm 200c}\sim 10^{14}h^{-1}M_\odot$ and decreases to 
$\sim 0.5$ at $M_{\rm 200c}\sim 3\times 10^{13}h^{-1}M_\odot$, and
hence the correction of the completeness due to $S(M)$ is relatively
minor. We find that the completeness after the $N_{\rm lim}$
correction is reasonably high ($> 0.9$) at high mass end, 
$M_{\rm 200c}\gtrsim 10^{14}h^{-1}M_\odot$, and decreases below
$10^{14}h^{-1}M_\odot$. This result is also broadly consistent with
the completeness estimated from X-ray clusters shown in
Figure~\ref{fig:compl}. A possible explanation of decreasing
completeness at low mass end is that the scatter in the richness-mass
relation increases rapidly toward lower richness, due to the larger
Poisson noise of the number of cluster member galaxies and more
significant contribution of background fluctuations to the richness
that enhances the scatter. In either case, the high completeness at
high cluster masses is encouraging from the viewpoint of using massive
optically-selected clusters as a cosmological probe.  

It is possible that the completeness changes with redshift. To check
the possible redshift dependence, we divide the mock halo sample into
two redshift bins with $0.3<z<0.6$ and $0.6<z<1.0$, and derive
completeness for both the subsamples. The result shown in
Figure~\ref{fig:mock_compl} indicates that the completeness indeed
depends on the redshift. For the low-redshift halos, the completeness
remains high down to $\sim 8\times 10^{13}h^{-1}M_\odot$, whereas for
high-redshift halos the completeness starts to drop at higher
masses. Decreasing completeness with increasing redshift for a fixed
halo mass may be partly due to decreasing red galaxy fraction which is
taken into account in the mock galaxy catalog construction (see
Appendix~3). In both redshift bins the completeness is close to unity
above $\sim 1.5 \times 10^{14}h^{-1}M_\odot$. This redshift evolution of the
completeness should also be taken into account for careful statistical
studies of CAMIRA clusters.

Again, we emphasize that our quantitative results from the mock
catalog depend on input models, such as cosmological parameters and
mass-richness relations for populating galaxies in halos, which can
differ from true values. For example, the redshift evolution of the
completeness discussed above depends crucially on the redshift
evolution of the input mass-richness relation and the input redshift
evolution of the red fraction in clusters (see  Appendix~3). In
particular, we note that our current mock catalog appears to
overestimate richness in the sense that the number density of
CAMIRA-detected clusters in the mock as a function of richness is
a factor of $\sim 4$ higher than in observations, which suggests
room for improvements of the mock. Nevertheless, the mock result
should still be useful for understanding qualitative behavior of the
completeness and purity of the CAMIRA algorithm, and both the mock
catalog analysis presented here and the comparison with other (such as
X-ray) cluster catalogs as presented in Section~\ref{sec:xray} are
important for characterizing the CAMIRA cluster catalog.

\section{Summary}\label{sec:summary}

We have presented a new optically-selected cluster catalog from the
HSC Survey. From the HSC Wide S16A dataset covering
$\sim 232$~deg$^2$, we have constructed a sample of 1921 clusters at
redshift $0.1<z_{\rm cl}<1.1$ 
and with richness $\hat{N}_{\rm mem}>15$. Based on the number density,
we infer that the rough mass limit of the cluster sample is $M_{\rm
  200m}\gtrsim 10^{14}h^{-1}M_\odot$. We have found that cluster
redshifts are accurate with the bias $\delta_z<0.005$ and the scatter
$\sigma_z<0.01$ for most of the redshift range. The photometric
redshift accuracy is comparable to that of optically-selected clusters
in SDSS, but HSC clusters extend to much higher redshifts. We have also
compared the HSC Wide S16A cluster catalog with X-ray clusters from
XXL and XMM-LSS surveys, finding tight correlations between richness
and X-ray properties. In addition, we have derived the distribution of
positional offsets of cluster centers using the X-ray clusters and
found that the fraction of well-centered clusters is $\sim 0.7$, with
no significant dependence on richness. We have analyzed mock galaxy
catalogs to study the completeness and purity. We have found a high
purity, and also the high completeness for halos with masses above
$\sim 10^{14}h^{-1}M_\odot$. The completeness depends on redshift such
that completeness for lower mass halos is higher at lower redshift.

This work presents the first cluster catalog from the HSC Survey, and
demonstrates the power of the HSC Survey for studies of high-redshift
clusters. The exquisite depth of the HSC Survey allows us to detect
almost all cluster member galaxies above $M_*\sim 10^{10.2}M_\odot$
even at $z\sim 1.1$, and hence allows a reliable cluster search at such
high redshifts. For instance, by extrapolating the result in this
paper, we expect to construct a large catalog of $\sim 1000$ clusters
with richness $\hat{N}_{\rm mem}>20$ at $z\sim 1$ from the final HSC
Wide Survey dataset covering $\sim 1400$~deg$^2$. We plan to calibrate
cluster masses using stacked weak lensing technique, as well as careful
comparisons with X-ray and SZ cluster catalogs, including an extended
XXL X-ray cluster catalog and SZ clusters from ACTPol \citep{niemack10}.

\begin{ack}
We thank Crist{\'o}bal Sif{\'o}n for useful comments, and an anonymous
referee for useful suggestions. 
This work is supported in part by World Premier International Research Center Initiative (WPI Initiative), MEXT, Japan, and JSPS KAKENHI Grant Number 26800093.
This work is in part supported by MEXT Grant-in-Aid for Scientific Research on Innovative 
Areas (No.~15H05887, 15H05892, 15H05893).
YTL acknowledges support from the Ministry of Science and Technology
grants MOST 104-2112-M-001-047 and MOST 105-2112-M-001-028-MY3.
HM is supported by the Jet Propulsion Laboratory, California Institute
of Technology, under a contract with the National Aeronautics and
Space Administration.

The Hyper Suprime-Cam (HSC) collaboration includes the astronomical communities of Japan and Taiwan, and Princeton University.  The HSC instrumentation and software were developed by the National Astronomical Observatory of Japan (NAOJ), the Kavli Institute for the Physics and Mathematics of the Universe (Kavli IPMU), the University of Tokyo, the High Energy Accelerator Research Organization (KEK), the Academia Sinica Institute for Astronomy and Astrophysics in Taiwan (ASIAA), and Princeton University.  Funding was contributed by the FIRST program from Japanese Cabinet Office, the Ministry of Education, Culture, Sports, Science and Technology (MEXT), the Japan Society for the Promotion of Science (JSPS),  Japan Science and Technology Agency  (JST),  the Toray Science  Foundation, NAOJ, Kavli IPMU, KEK, ASIAA,  and Princeton University.

The Pan-STARRS1 Surveys (PS1) have been made possible through contributions of the Institute for Astronomy, the University of Hawaii, the Pan-STARRS Project Office, the Max-Planck Society and its participating institutes, the Max Planck Institute for Astronomy, Heidelberg and the Max Planck Institute for Extraterrestrial Physics, Garching, The Johns Hopkins University, Durham University, the University of Edinburgh, Queen's University Belfast, the Harvard-Smithsonian Center for Astrophysics, the Las Cumbres Observatory Global Telescope Network Incorporated, the National Central University of Taiwan, the Space Telescope Science Institute, the National Aeronautics and Space Administration under Grant No. NNX08AR22G issued through the Planetary Science Division of the NASA Science Mission Directorate, the National Science Foundation under Grant No. AST-1238877, the University of Maryland, and Eotvos Lorand University (ELTE).
 
This paper makes use of software developed for the Large Synoptic Survey Telescope. We thank the LSST Project for making their code available as free software at http://dm.lsst.org.

Based [in part] on data collected at the Subaru Telescope and retrieved from the HSC data archive system, which is operated by the Subaru Telescope and Astronomy Data Center at National Astronomical Observatory of Japan.
\end{ack}

\section*{Appendix 1. Cluster catalogs}
We construct three cluster catalogs from the HSC S16A data. The
first catalog is from the Wide region without applying for a bright
star mask. This catalog is used in the analysis throughout the
paper. The second catalog is also from the Wide region, but with the
bright star mask. The bright star mask removes object near bright
stars, and the current mask is designed conservatively such that it
removes nearly $\sim 10\%$ of objects. 
The current cluster catalog with the star mask suffers
  from the effect of masking bright galaxies, which is discussed in
  Appendix 2.
The third catalog is from the
Deep region covering $\sim 25$~deg$^2$. Even though the Deep data are
deeper than the Wide data, we use the same selection criteria for
constructing an input galaxy catalog, including the conservative
$z$-band magnitude limit of $z<24$. These catalogs are shown in
Supplementary Tables 1, 2, and 3.

\section*{Appendix 2. Effect of the bright star mask on cluster finding}
In addition to overly conservative sizes of mask regions, a known
issue of the current version of the bright star mask is that it also
masks bright nearby galaxies \citep{aihara17}. This is because a bright
object catalog from the Naval Observatory Merged Astrometric Dataset
\cite[NOMAD;][]{zacharias05}, which we use to select mask regions,
often misinterprets nearby bright galaxies as stars. We check the
impact of masking bright galaxies on cluster finding by comparing
cluster catalogs with and without star masks.

\begin{figure}
 \begin{center}
  \includegraphics[width=8cm]{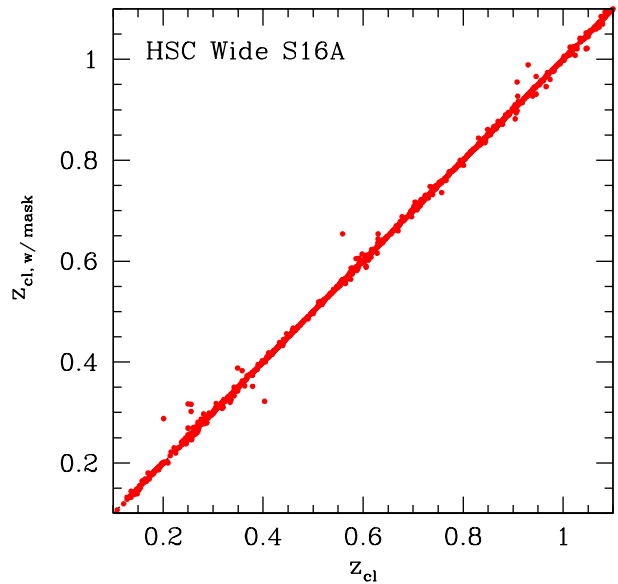} 
 \end{center}
 \caption{
 The comparison of cluster photometric redshifts $z_{\rm cl}$ between HSC Wide S16A
 cluster catalogs with and without the bright star mask. 
}\label{fig:comp_zsm} 
\end{figure}

\begin{figure}
 \begin{center}
  \includegraphics[width=8cm]{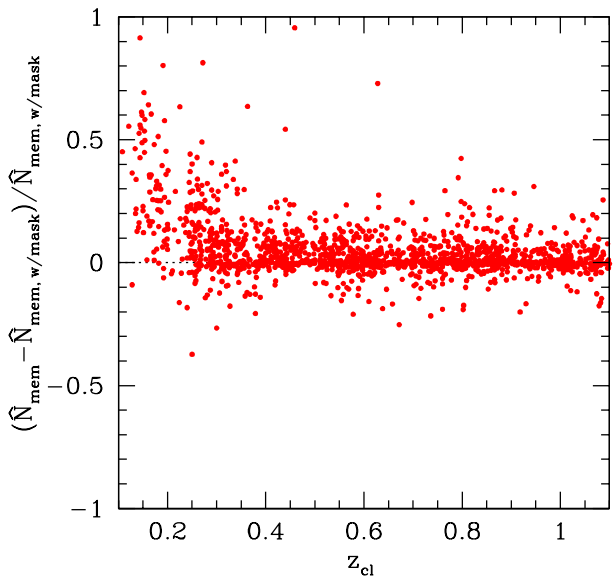} 
 \end{center}
\caption{ 
The comparison of richness $\hat{N}_{\rm mem}$ between HSC Wide S16A
 cluster catalogs with and without the bright star mask. 
 }\label{fig:check_sm_dnz} 
\end{figure}

We match HSC Wide S16A cluster catalogs with and without the bright
star mask in the same manner as in Section~\ref{sec:sdss}. The
comparison of cluster photometric redshifts shown in
Figure~\ref{fig:comp_zsm} indicates that photometric redshifts agree
well between these two cluster catalogs. In contrast, the comparison
of richness in Figure~\ref{fig:check_sm_dnz} shows a clear bias at low
redshifts, $z\lesssim 0.4$, such that richness in the cluster catalog
with the bright star mask is systematically lower than in the cluster
catalog without the bright star mask. We find that this is due to
masking of bright galaxies as mentioned above, which masks red member
galaxies near BCGs, as well as BCGs themselves, leading to smaller
richness estimates for those clusters. Therefore, for any cluster
studies that involve low-redshift clusters, we recommend not to use 
the cluster catalog with the bright star mask. For studies of
high-redshift ($z\gtrsim 0.4$) clusters, however, the effect of
masking bright galaxies is negligible and we expect we can safely use
the cluster cluster with the bright star mask.

As discussed in \citet{aihara17}, we are working on revising the
bright star mask, and we expect that this issue will be resolved in
future versions of HSC CAMIRA cluster catalogs.

\section*{Appendix 3. Description of the mock galaxy catalog}
The set of mock galaxy catalogs we use for testing the cluster finding
algorithm is produced by combining a ``field'' galaxy population with a
``cluster'' galaxy population.  In a nutshell, the ``field''
population is generated from well-studied extragalactic fields with
rich multiwavelength data (particularly in X-rays), so that galaxies
associated with massive galactic systems such as clusters and groups
can be removed (leaving only the field galaxies). On the other hand,
the ``cluster'' population is obtained by populating dark matter halos
with galaxies from real, observed clusters.  Below we briefly describe
how each of these is produced.  

For the field galaxy catalog, we use the HSC Subaru Strategic Program
observation of the COSMOS \citep{scoville07} field as the parent
sample.  More specifically, we adopt the Wide-layer depth $grizy$-band 
photometry of the Ultradeep-layer observation.  We further remove
galaxies that are likely associated with known structures in the
COSMOS field, such as galaxy groups and clusters, by excluding $\sim
4000$ galaxies that have membership probability $P_{\rm mem}>0.5$
according to the galaxy group catalog of \citet{george11}. 

For generating the cluster galaxy catalog, our first step is to extract
halo catalogs from a lightcone constructed from a large $N$-body
simulation (see \citealt{sehgal10} for more details of the simulation
and the lightcone creation).
The halos are extracted from the $N$-body simulation
which has a box size of $1000\,h^{-1}$Mpc, using a friends-of-friends
algorithm with a linking length of 0.2 mean particle separation.
In summary, the lightcone is complete for
halos with $M_{\rm 200c}\ge 10^{13}M_\odot$ and $z\le 3$.  We consider a patch
in the lightcone that is of the same geometry and area as the COSMOS
field, and only include halos at $z<1.2$. We then populate these halos with
galaxies that are probable members of real clusters and groups
observed by the HSC, following a galaxy number--halo mass relation  
\begin{equation}
N(M,z) = A \left( \frac{M_{\rm 200c}}{10^{14} M_\odot} \right)^{b} (1+z)^{c}
\label{eq:hon}
\end{equation}
with $A=47$, $b=0.85$, $c=-0.1$, which is consistent with observations
of \citet{lin04,lin06} for galaxies more luminous than $M^\star(z)+4$,
where $M^\star$ is the evolving characteristic magnitude of the galaxy
luminosity function in $z$-band. We note that the definition of this
input $N$ differs from the richness in CAMIRA. The input richness is
defined for all cluster member galaxies rather than red member
galaxies used in CAMIRA, and also has a different luminosity
threshold. Given a halo mass $M_{\rm 200c}$, we draw a Poisson random
number with the mean given by the above expression as the total number
$N$ of galaxies to be assigned to the halo.  The partition of red and
blue galaxies is specified by a red fraction function, following the
fitting formula of \citet{hansen09},  
\begin{equation}
f_{\rm red}(N,z) = g(z) {\rm erf}\left[\log N - \log h(z)\right] +0.69,
\end{equation}
where $g(z)=0.206-0.371z$, $h(z)=-3.6+25.8z$.
 We note that the fitting formula of \citet{hansen09} was defined in 
 the redshift range $0.1\leq z\leq 0.3$, and we extrapolate this result
 to $z=1.2$, mainly because of the lack of detailed quantitative
 studies of red fractions at high redshifts. Since the redshift
 evolution of the completeness and purity estimated from mock
 catalogs crucially depend on the input redshift evolution of the red
 fraction, our mock result should be taken with caution. 
The numbers of red and blue galaxies of the halo in question are
$N_{\rm red} =  f_{\rm red} N$ and $N_{\rm blue}=(1-f_{\rm red})N$,
respectively.  The rationale behind the adaptation of a red fraction
function is that, due to projection effects, the blue-cloud galaxies
usually outnumber that of red-sequence ones, and thus if one simply
draws in random $N$ galaxies the color distribution will be strongly
biased and uncharacteristic of cluster galaxies. 

\begin{figure}
 \begin{center}
  \includegraphics[width=8cm]{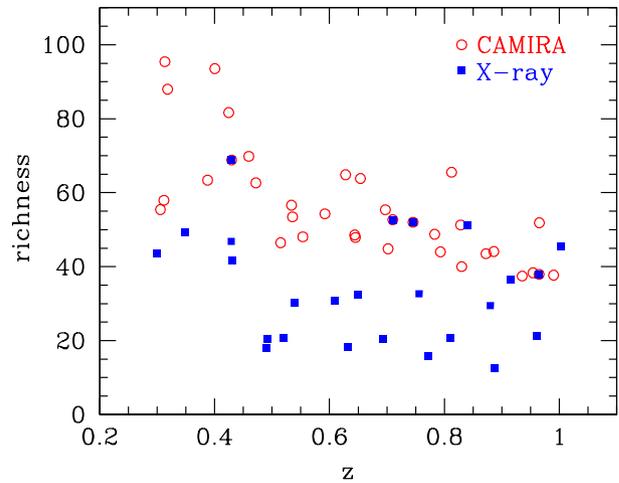} 
 \end{center}
 \caption{
Comparison of the clusters used to ``populate'' the halos
between our default set (open circles; 35 clusters) and the test set
(filled square; 25 clusters).  The default set uses richest CAMIRA
clusters, while the test set employs those CAMIRA clusters that are
also detected in the X-rays, which are mostly from the XXL and XMM
Cluster Survey \cite[XCS;][]{mehrtens12}.
}\label{fig:mock_test} 
\end{figure}

\begin{figure}
 \begin{center}
  \includegraphics[width=8cm]{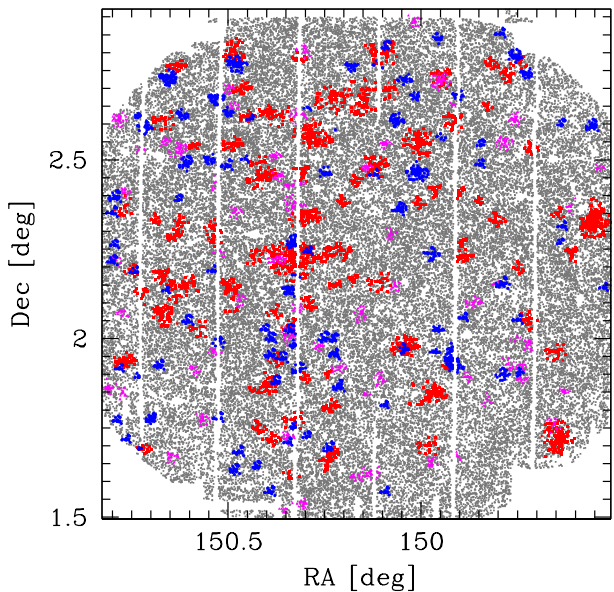} 
 \end{center}
\caption{An example of our mock galaxy catalogs for testing cluster
  finding algorithms. Small gray dots show the distribution of
  ``field'' galaxies, taken from real HSC observations. Red squares,
  magenta triangles, and blue circles show the distribution of
  ``cluster'' galaxies at redshift $0.3<z<0.6$, $0.6<z<0.9$, and
  $0.9<z<1.2$, respectively. 
}\label{fig:mock_demo} 
\end{figure}

As for the actual cluster member galaxy data used in this procedure,
we have selected the top 5 richest clusters detected by CAMIRA in each
of the redshift bins from $z=0.3$ to $z=1.0$ with bin width of
0.1. For each of these real clusters, we initially consider all galaxies
lying within a projected distance of $\tilde{R}_{\rm 200c}$.  Here
$\tilde{R}_{\rm 200c}$ is a rough estimate of the true value for these
massive clusters, obtained by taking average of 5 richest mock
clusters from the public MICE mock catalog \citep{carretero15}
selected in the same redshift range, over an area of
200~deg$^2$\footnote{Once the richness—mass relation is calibrated
  via weak lensing from the HSC survey itself, we will use it for the
  initial estimate of $R_{200}$, thus bypassing this step.}.
The subsequent treatment is different for red and blue galaxies.  For the
galaxies on the red sequence, we simply select the needed 
$N_{\rm red}$ galaxies, giving preference to the brighter galaxies,
using simple red galaxy selections in the color-magnitude diagram.
A caveat is that this procedure of populating red
  galaxies from brighter to fainter galaxies using the richest
  clusters may introduce bias in mock galaxy population, such as the
  overestimate of the bright end of the low-mass clusters luminosity
  functions, which is ignored here.
For the blue galaxies, we use photometric redshifts \citep[using
  EAZY;][]{brammer08} derived from the HSC Survey data to
facilitate the selection, taking into account the accuracy of the
photometric redshift as a function of magnitude, and 
draw in random $N_{\rm blue}$ galaxies from the blue cloud by giving
higher priority to brighter galaxies.  For a 
given halo in the lightcone, we select the real cluster that is
closest in terms of both redshift and richness.  To account for the
fact that real clusters do not lie at the same redshifts of the halos,
we use a passively evolving stellar population model based on the
\citet{bruzual03} model to shift the real galaxies either forward or
backward in time with a typical magnitude shift of 0.03. Finally,
while we keep the relative orientation of galaxies with respect to the
cluster center (defined to be the location of the brightest cluster
galaxy), we rescale the radial distance by the halo mass, assuming an
NFW profile \citep{navarro97} with the concentration parameter of $c=5$. 
The value of the concentration parameter we adopted is
a typical value for the dark matter distribution of halos with masses
and redshifts of our interest \citep[e.g.,][]{gao08}.

In principle, we could include member galaxies in
clusters selected by other means, such as X-ray or SZ effect.  For the
current version of the mock, we have opted to use CAMIRA-selected
clusters as these are detected over much larger area than existing
X-ray surveys, and thus represent currently the best available massive
cluster sample. As an exploratory example, we also attempt another set
of mocks (hereafter referred to as the ``test set''), which differs
from our default set in the source of input clusters. Instead of using
the richest CAMIRA clusters, we resort to known X-ray clusters covered
by the current HSC footprint that are also matched to the CAMIRA
cluster catalog.  As mentioned above, one important limitation for
adopting the X-ray-selected clusters is that due to the small area
coverage in the X-ray surveys (only a small fraction of HSC survey
footprint has deep X-ray data for detecting clusters out to high
redshifts), the clusters are typically much poorer than those used in
our default mock set (see Figure~\ref{fig:mock_test}), 
which makes our galaxy selection process more vulnerable to
contamination from foreground/background galaxies and may lead to
lower completeness. In the future we should use X-ray or SZ selected
clusters in order to avoid any selection bias of optically selected
clusters which by definition contains prominent concentrations of
red cluster member galaxies.

After all the halos in the lightcone patch have been populated with
galaxies from real galactic systems (modulo shifts in redshift using
passive evolution), we finally remove those galaxies lying in the
gap and bright star regions seen in the field galaxy catalog mentioned
above, and combine the resulting galaxy catalog with the field galaxy
catalog.  On average, on top of the $\sim 130,000$ galaxies in the
field, we add about 7000 cluster galaxies. We show an example of
the distribution of mock galaxies in Figure~\ref{fig:mock_demo}. We
note that the spatial distribution of our mock galaxy sample contains
several vertical stripes of gaps, which are not seen in most of the
HSC Wide layer data. Since these stripes are expected to reduce the
completeness estimated from these mock galaxy samples, we expect that
our mock analysis results presented in Section~\ref{sec:mock} are
conservative. 

As our $N$-body lightcone covers an octant of the sky, we can extract
many patches of the mock sky identical in geometry to the COSMOS
field, and thus rapidly generate mock catalogs with area of hundreds
of square degrees or more, although the spatial distribution of the
field galaxies would be repeated and discontinuous if all the patches
are combined together. Furthermore, as the same set of large scale
structure in COSMOS are repeated in all the mocks, the effects of
large scale structure in cluster finding may not be fully captured in
this approach. We do note that, halos found in filamentary structures
surrounding massive halos are also populated with galaxies, and
therefore in our mocks the projection effects from surrounding
large-scale structures are accounted for to some degree.
Finally, it should be emphasized that the way we create the cluster
galaxies inevitably rely on several assumptions on the redshift
evolution of the cluster galaxy population (e.g., the evolution of the
halo occupation number and the red galaxy fraction), and thus our
mocks should be regarded as a model, which still need to be calibrated
against detailed observations of distant clusters
\citep[e.g.,][]{hennig17}.  A proper way to obtain completeness and
purity from our mocks is thus to generate sets of mocks that cover the
ranges of plausible values of model parameters (e.g., $c$ in
equation~\ref{eq:hon}, form and normalization of $f_{\rm red}$), and
marginalize over these mocks. 


\end{document}